\documentclass[12pt]{article}

\usepackage{graphicx,psfrag,epsf}
\usepackage{enumerate}
\usepackage[sort,numbers]{natbib}
\usepackage{mathtools}
\usepackage{booktabs}
\usepackage{color}
\usepackage[english]{babel} 
\usepackage[protrusion=true,expansion=true]{microtype} 
\usepackage{amsmath,amsfonts,amsthm}
\usepackage{dsfont}
\usepackage{amssymb}
\usepackage{chngcntr}
\usepackage[toc,page]{appendix}
\usepackage{textcomp}
\usepackage{url}
\usepackage{hyperref}

\usepackage{array}
\usepackage{booktabs} 
\usepackage{setspace}
\usepackage{amsmath}
\usepackage{xcolor}
\usepackage{soul}
\usepackage{multirow}
\usepackage{enumitem}
\usepackage{microtype} 
\usepackage[font = small,labelfont=bf,textfont=it]{subcaption}
\usepackage{xcolor}
\usepackage{tabularx}
\usepackage[font = small,labelfont=bf,textfont=it]{caption} 
\usepackage{footnote}
\usepackage{algorithm}
\usepackage[noend]{algpseudocode}
\usepackage{etoolbox}
\usepackage{tabularx}
\usepackage{authblk}
\usepackage{caption}
\usepackage{indentfirst}
\usepackage{enumitem}
\newlist{steps}{enumerate}{1}
\setlist[steps, 1]{label = Step \arabic*:}

\algblock{ParFor}{EndParFor}
\algnewcommand\algorithmicparfor{\textbf{for}}
\algnewcommand\algorithmicpardo{\textbf{do\ parallel}}
\algnewcommand\algorithmicendparfor{\textbf{end\ parallel\ for}}
\algrenewtext{ParFor}[1]{\algorithmicparfor\ #1\ \algorithmicpardo}
\algrenewtext{EndParFor}{\algorithmicendparfor}

\makeatletter
\def\BState{\State\hskip-\ALG@thistlm}
\DeclareMathOperator*{\argmax}{\arg\!\max}
\newcommand{\distas}[1]{\mathbin{\overset{#1}{\kern\z@\sim}}}%
\newcommand{\bm}[1]{\mathbf{#1}}
\newcommand{\bs}[1]{\boldsymbol{#1}}
\newsavebox{\mybox}\newsavebox{\mysim}
\newcommand{\distras}[1]{%
  \savebox{\mybox}{\hbox{\kern3pt$\scriptstyle#1$\kern3pt}}%
  \savebox{\mysim}{\hbox{$\sim$}}%
  \mathbin{\overset{#1}{\kern\z@\resizebox{\wd\mybox}{\ht\mysim}{$\sim$}}}%
}

\newcommand{\be}{\begin{equation}}
\newcommand{\ee}{\end{equation}}
    \newcommand{\bi}{\begin{itemize}}
\newcommand{\ei}{\end{itemize}}
\newcommand{\ben}{\begin{enumerate}}
\newcommand{\een}{\end{enumerate}}

\newcolumntype{K}[1]{\geq {\centering\arraybackslash}p{#1}}
\allowdisplaybreaks

\makeatother

\let\oldbibliography\thebibliography
\renewcommand{\thebibliography}[1]{\oldbibliography{#1}
\setlength{\itemsep}{0pt}} 

\newcommand{\blind}{1}

\patchcmd{\footnotemark}{\stepcounter{footnote}}{\refstepcounter{footnote}}{}{}
\setcitestyle{square}

\begin{document}

\def\spacingset#1{\renewcommand{\baselinestretch}%
{#1}\small\normalsize} \spacingset{1}

\if1\blind
{
  \title{\bf Additive Multi-Index Gaussian process modeling, with application to multi-physics surrogate modeling of the quark-gluon plasma}
  \small
   \author{Kevin Li\footnote{Department of Statistical Science, Duke University},\; Simon Mak$^*$, J.-F. Paquet\footnote{Department of Physics and Astronomy \& Department of Mathematics, Vanderbilt University},\; Steffen A. Bass\footnote{Department of Physics, Duke University}}
  \maketitle
} \fi


\if0\blind
{
  \bigskip
  \bigskip
  \bigskip
  \begin{center}
    {\LARGE\bf Additive Multi-Index Gaussian process modeling, with application to multi-physics surrogate modeling of heavy-ion collisions}
\end{center}

  \medskip
} \fi

\begin{abstract}
The Quark-Gluon Plasma (QGP) is a unique phase of nuclear matter, theorized to have filled the Universe shortly after the Big Bang. A critical challenge in studying the QGP is that, to reconcile experimental observables with theoretical parameters, one requires many simulation runs of a complex physics model over a high-dimensional parameter space. Each run is computationally very expensive, requiring thousands of CPU hours, thus limiting physicists to only several hundred runs. Given limited training data for high-dimensional prediction, existing surrogate models often yield poor predictions with high predictive uncertainties, leading to imprecise scientific findings. To address this, we propose a new Additive Multi-Index Gaussian process (AdMIn-GP) model, which leverages a flexible additive structure on low-dimensional embeddings of the parameter space. This is guided by prior scientific knowledge that the QGP is dominated by multiple distinct physical phenomena (i.e., multi-physics), each involving a small number of latent parameters. The AdMIn-GP models for such embedded structures within a flexible Bayesian nonparametric framework, which facilitates efficient model fitting via a carefully constructed variational inference approach with inducing points. We show the effectiveness of the AdMIn-GP via a suite of numerical experiments and our QGP application, where we demonstrate considerably improved surrogate modeling performance over existing models. 



\end{abstract}

\vfill

\newpage
\spacingset{1.5} 

\section{Introduction} \label{sec:intro}

The Quark-Gluon Plasma (QGP) is an exotic phase of nuclear matter whose constituents, namely quarks and gluons, are the elementary building blocks of protons, neutrons and nuclei. The QGP is theorized to have filled the Universe shortly after the Big Bang, and the study of this plasma sheds light on the conditions present in the early Universe. Recent promising work \cite{everett2021multisystem,everett2021phenomenological,liyanage2022efficient,ji2021graphical} has focused on the use of virtual simulations from complex physics models, to reconcile physical parameters $\bm{x} \in \mathbb{R}^d$ with experimental data from particle colliders. Despite this progress, there remains a critical bottleneck: such analysis requires many runs from the expensive simulator $f(\cdot)$ at different parameters $\bm{x}$, each requiring \textit{thousands} of CPU hours \cite{cao2021determining}. This is exacerbated by the relatively high dimension of $\bm{x}$, which may be on the order of 20 parameters for full-scale QGP studies. Given a limited computing budget, one can afford only hundreds of simulation runs over the high-dimensional parameter space, which results in imprecise scientific findings \citep{liyanage2022efficient}.

\begin{figure}[!t]
    \centering
    \includegraphics[width=0.9\textwidth]{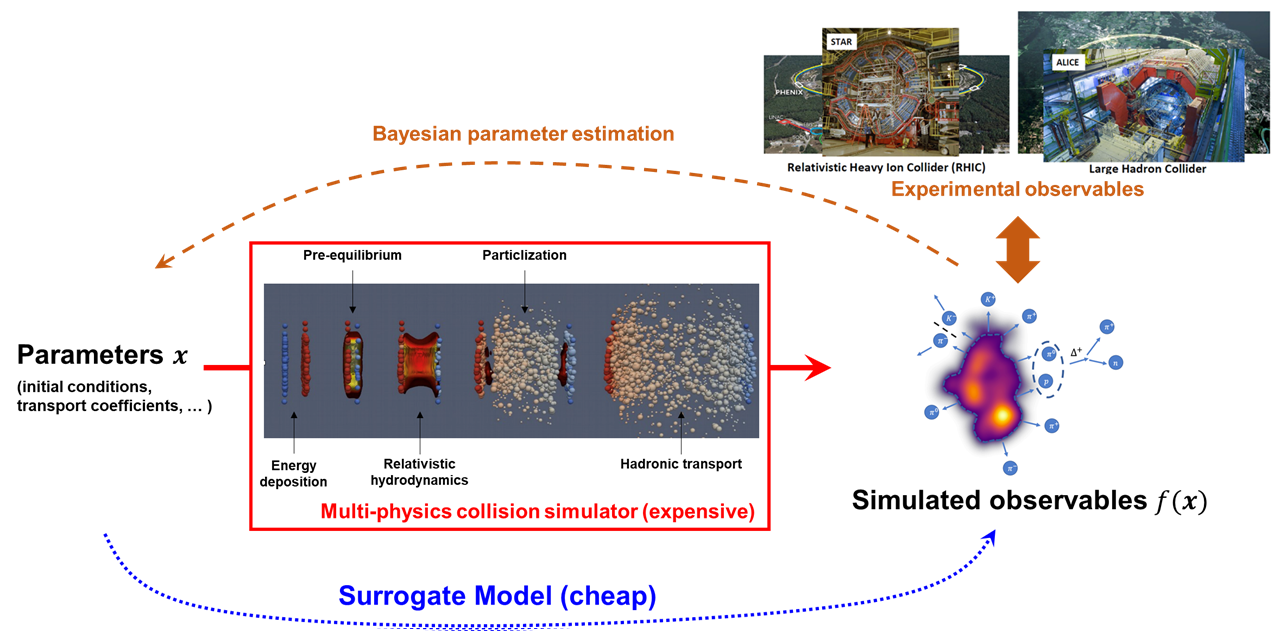}
    \caption{Visualizing the role of probabilistic surrogate modeling for studying the quark-gluon plasma using expensive multi-physics simulations (figure adapted from images from the MADAI collaboration, \cite{everett_thesis}, Brookhaven National Laboratory and CERN).}
    \label{fig:vis}
\end{figure}

Surrogate modeling \cite{santner2003design, gramacy_surrogates_2020} provides a promising solution; see Figure \ref{fig:vis} for a visualization. The idea is simple but effective: simulation experiments are first performed at carefully selected parameter points, then used as training data to fit an \textit{surrogate model} that efficiently \textit{emulates} and quantifies uncertainty on the simulation response surface $f(\cdot)$. There is a rich and growing literature on probabilistic surrogate modeling, particularly using Gaussian processes (GPs; \cite{stein1999interpolation,gramacy_surrogates_2020}), which offer a flexible Bayesian framework for efficient prediction and uncertainty quantification (UQ). This includes the seminal works on GP-based surrogates \cite{sacks1989design,currin1991bayesian,WELCH1992}, as well as recent extensions of such models for complex physics and engineering applications \cite{ji2022multi,chen2020function, sauer_gp, chili}. However, such models are known to suffer from a curse-of-dimensionality; when the number of parameters $d$ becomes large, one may require a sample size $n$ growing exponentially in $d$ to achieve satisfactory predictive performance \cite{van2008rates}, which can easily become prohibitively expensive. Given a tight budget on run size $n$, such models can thus yield poor predictions with unacceptably high uncertainties in high dimensions.


One way to address this is to learn plausible low-dimensional structures in $f$ and integrate such structures for predictive modeling. There has been recent work in this direction for Bayesian surrogate modeling, particularly using Gaussian processes. \cite{linkletter2006variable} explored the use of variable selection within GPs for identifying sparsity (see also \cite{ben_salem_sequential_2019} and \cite{eriksson_high-dimensional_2021}). \cite{gramacy2012gaussian} proposed single-index GP models, which can identify and quantify uncertainty on an active one-dimensional linear embedding of the parameter space. \cite{snelson2012variable, seshadri_dimension_2019, bilionis2016gaussian} further extended these models to identify higher-dimensional linear embeddings within GPs. Such extensions, however, largely cannot quantify uncertainties in estimating the underlying embedding, which as we show later may result in highly overconfident predictions. \cite{wang_bayesian_2016,letham_re-examining_2020} investigated the use of random embeddings within GPs. There is also a rich literature on leveraging additive components of low-dimensional functions within GPs; see \cite{duvenaud_additive_2011, durrande_additive_2012, pmlr-v37-kandasamy15, gilboa_scaling_2013, delbridge_randomly_2020}.

Despite this literature, there are two critical limitations of existing models for our application. First, these approaches do not directly elicit (and thus model for) the specific embedded low-dimensional structure guided by the \textit{multi-physics} nature of the physical system. Hearkening back to the well-known Buckingham-$\pi$ theorem \cite{buckingham1914physically}, it is accepted that complex systems have structured low-dimensional manifold embeddings that represent a sparse number of dominant physics. For the QGP, the simulated collision observables are known to capture \textit{several} types of dominant physics (e.g., {geometry and quantum fluctuations, relativistic hydrodynamic expansion}), \textit{each} of which depends on a \textit{low-dimensional embedding} of the parameter space. This notion of {multi-physics}, where the simulator integrates multiple distinct types of physics, is widely used for simulating complex phenomena, from reacting flows \cite{mak2018efficient} to particle collisions \cite{kumar2023inclusive}. However, existing surrogate models do not elicit nor model for this embedded structure, and thus may yield poor predictions with high uncertainties when such multi-physics are present (as we show later in Section \ref{sec:limitations}). Second, existing models largely do not account for \textit{uncertainties} in estimating the underlying embedded structure (more on this in Section \ref{sec:gp}). Particularly with a limited sample size $n$ in high dimensions $d$, neglecting such uncertainties results in a wildly overconfident surrogate model, which can easily lead to spurious scientific conclusions.

To address these limitations, we propose a novel Additive Multi-Index Gaussian process (AdMIn-GP) model, which integrates the elicited low-dimensional embedded structures from multi-physics for surrogate modeling. The AdMIn-GP features a flexible additive model of GPs with each component active on different low-dimensional linear embeddings of the parameter space, to model the presence of multiple dominant physics within the complex simulation system. We present a novel variational inference approach for efficient and probabilistic prediction from the AdMIn-GP, leveraging carefully-constructed shrinkage priors on the embedding matrices. We demonstrate the improved predictive and uncertainty quantification performance of the AdMIn-GP over existing models in a suite of numerical experiments and our motivating high-energy physics application, thus showing that when multi-physics is present, the integration of such structure can greatly improve surrogate modeling with limited data. We then highlight how the AdMIn-GP sheds light on the extraction and interpretation of dominant multi-physics in the quark-gluon plasma, which can help guide scientific discoveries.

This paper is organized as follows. Section \ref{sec:QGP} describes the motivating quark-gluon plasma application, and outlines key limitations of existing surrogate models for this problem. Section \ref{sec:vmegp} introduces the AdMIn-GP model, and presents an efficient variational inference algorithm for model training and selection. Section \ref{sec:exp} compares the AdMIn-GP to the state-of-the-art in a suite of numerical experiments. Section \ref{sec:app} then applies the AdMIn-GP for our application on the surrogate modeling of the QGP, and discusses its potential for guiding extraction of dominant multi-physics. Section \ref{sec:con} concludes the paper.

\section{The quark-gluon plasma}
\label{sec:QGP}
We first outline key challenges underlying the surrogate modeling of the QGP, then survey existing methods and investigate their limitations for our application.

\subsection{Simulating the QGP via heavy-ion collisions}
\label{sec:qgpe}

As the theorized form of matter permeating the early Universe a few microseconds after the Big Bang, the QGP is an important topic of study in nuclear physics. This plasma is produced and explored in physical experiments via the collisions of heavy nuclei at velocities close to the speed of light. Such experiments are conducted at large particle colliders at Brookhaven National Laboratory and the European Organization for Nuclear Research (CERN). The temperature and pressure of these collisions converts the colliding nuclei into a plasma of subatomic particles, namely, quarks and gluons. Due to its rapid hydrodynamic expansion, the plasma quickly cools down and decays. The decay products of the quark-gluon-plasma can subsequently be observed by particle detectors.


To study properties of the QGP, one requires the coupling of data taken by these particle detectors with complex nuclear physics \textit{simulation} models, to reconcile plasma properties (denoted by parameters $\bm{x} \in \mathbb{R}^d$) with experimental data. These nuclear collision simulations are typically performed in several successive stages (following \cite{everett2021multisystem}) to faithfully capture the complex multi-physics phenomena. The first stage is the impact of the colliding nuclei, and the deposition of energy from this impact. This is followed by a pre-equilibrium phase, where the system of quarks and gluons approaches local equilibrium. The third phase simulates the evolution of plasma utilizing relativistic viscous fluid dynamics, where the strongly-interacting plasma expands and rapidly cools down. The fourth stage models particlization -- the transition from the strongly-coupled quark-gluon plasma to individual bound states of quarks and gluons (called ``hadrons''), whose interaction strength is insufficient to maintain a cohesive hydrodynamic expansion. The final stage simulates the interaction and subsequent decay of the hadronic bound states, until their density is so low that interactions cease and the particles travel to the particle detectors. The behavior of this collision system is controlled by parameters $\bm{x}$, which characterize the energy deposition at the initial nuclear impact, its evolution towards local equilibrium, the transport coefficients of the QGP (e.g., shear and bulk viscosity), and the recombination of quarks and gluons into normal nuclear bound states; details can be found in \cite{everett2021multisystem}. This complex simulator thus captures the interplay between \textit{multiple} types of physics, including finite density quantum field theory, relativistic viscous hydrodynamics, relativistic transport, and electromagnetic interactions.

Given parameters $\bm{x}$, we obtain a virtual simulation of an observable $f(\bm{x})$ that one might observe from the decay of the plasma. The goal is then to find parameters that best match the experimental observables from particle colliders. In recent years, Bayesian inference with such simulators have led to a deeper understanding of the properties of ultradense nuclear matter
\cite{everett2021phenomenological,everett2021multisystem}. The computational cost of a full-scale study of nuclear collisions is considerable, however, requiring \textit{thousands} of CPU hours per parameter. To address this critical bottleneck, surrogate models are increasingly used to efficiently emulate the expensive simulator \cite{petersen2011constraining,novak2014determining,bernhard2019bayesian,everett2021multisystem,liyanage2022efficient}; see Figure \ref{fig:vis} for a visualization. Of course, its success hinges on the ability to fit an accurate surrogate model with low predictive uncertainty. This is a \textit{highly challenging} task for the QGP: the parameter space for $\bm{x}$ is quite high-dimensional ($d=17$), and \textit{many} training design points are thus needed to sufficiently populate this high-dimensional space. However, each design point necessitates an \textit{expensive} run of the simulator $f(\cdot)$, which requires thousands of CPU hours. Given the computational budget for our project, this limits us to $n\approx500$ design points. With such limited data, the careful construction of the surrogate model is paramount for cost-efficient scientific discovery. 

\subsection{Existing GP surrogate models}
\label{sec:gp}
Much of the literature on probabilistic surrogate modeling involves Gaussian process models. We first provide below a brief review of GP modeling (see \cite{santner2003design} for further details), then discuss existing GP models that may apply for our problem.

Let $f: \mathbb{R}^d \rightarrow  \mathbb{R}$ denote the black-box function representing the observable simulated from the expensive simulator model. Here, the simulated plasma observables are known to be noisy, with noise corruption well-modeled via homoskedastic Gaussian noise (see, e.g., \cite{everett2021multisystem}). Given simulated parameters $\bm{x}_1, \cdots, \bm{x}_n$, we assume the simulated observables follow:
\begin{equation}
y_i = f(\bm{x}_i) + \epsilon_i, \quad \epsilon_i \distas{i.i.d.} \mathcal{N}(0,\sigma^2), \quad i = 1, \cdots, n.
\label{eq:obs}
\end{equation}

We then adopt a GP prior on the unknown response surface $f$, given as:
\begin{equation}
f(\cdot) \sim \text{GP}\{\mu,k(\cdot,\cdot)\}.
\label{eq:gp}
\end{equation}
Here, $\mu$ is a mean parameter for the stochastic process, and $k(\cdot,\cdot)$ is its covariance function that dictates smoothness of sample paths. Given a lack of prior knowledge, $k(\cdot,\cdot)$ can be taken as the squared-exponential or Mat\'ern kernel \cite{santner2003design}. Conditional on simulated data $\bm{y} = (y_1, \cdots, y_n)$, the posterior predictive distribution of $f$ at a new parameter $\bm{x}^*$ is:
\begin{align}
\small
\begin{split}
f(\bm{x}^{*})&|y_1, \cdots, y_n \sim \mathcal{N}\{\mu_n(\bm{x}^{*}), \sigma^2_n(\bm{x}^{*})\},\\
\mu_n(\bm{x}^{*}) &= \mu+ \bm{k}_{*,n}^T(\bm{K}_{n,n} + \beta^{-1} \bm{I}_{n \times n})^{-1} (\bm{y} - \bs{\mu}), \; \sigma^2_n(\bm{x}^{*}) = k(\bm{x}^{*}, \bm{x}^{*}) - \bm{k}_{*, n}^T(\bm{K}_{n,n} + \beta^{-1}\bm{I}_{n\times n})^{-1} \bm{k}_{n, *}.
\label{eq:gpeqns}
\end{split}
\end{align}
Here, $\bm{K}_{n,n} = [k(\bm{x}_i, \bm{x}_j)]_{i,j=1}^n$ is the covariance matrix for the design points, and $\bm{k}_{*, n}  = [k(\bm{x}^{*}, \bm{x}_i)]_{i=1}^n$ is the covariance vector between the design points and the new parameter. Equation \eqref{eq:gpeqns} highlights a key advantage of GP surrogates: it provides closed-form expressions for prediction (emulation) and uncertainty quantification, which can greatly speed-up downstream uses of the surrogate for optimization \cite{jones1998efficient,chen2019hierarchical} and design \cite{Binois_rep_exp_2018}.




One limitation of standard GP surrogates, as mentioned earlier, is that it suffers from a \textit{curse-of-dimensionality}: as the number of parameters $d$ grows large, one may require a sample size $n$ growing exponentially in $d$ to achieve satisfactory predictions (see, e.g., minimax and posterior contraction rates for GP models in \cite{van2000empirical,van2008rates}). For the current application with $d=17$ parameters and highly expensive simulation runs, such a sample size is unachievable with any reasonable budget! Given limited runs, one must then be judicious in identifying low-dimensional structure for improving predictions. We review existing literature on this below.

An important early work in this direction is \cite{gramacy2012gaussian}, which proposed a fully Bayesian single-index GP model (SIM-GP); this is further extended in \cite{hu2013bayesian} for quantile regression. The key idea is in representing $f$ as $f(\bm{x}) = g(\bm{m}^T \bm{x})$, where $c = \bm{m}^T \bm{x}$ represents a \textit{single} latent variable that accounts for variation over the response surface, and $g$ follows the aforementioned GP model. The loading vector $\mathbf{m}$ is then inferred in a fully Bayesian fashion via Markov chain Monte Carlo, using appropriate priors on $\bm{m}$ and  GP length-scale parameters. However, as noted in \cite{snelson2012variable}, the SIM-GP and its variants encounters difficulties when the embedded structures in $f$ are more complex and have dimensionality greater than one; we shall see this later in our application.

There has been much recent work on exploring a more flexible low-dimensional embedding via \textit{active subspaces}. Here, $f$ is modeled as the form $f(\bm{x}) = g(\bm{M}\bm{x})$, where $\bm{M} \in \mathbb{R}^{p \times d}$, $p \ll d$, is a matrix that maps the original parameters $\bm{x}$ onto an active lower-dimensional latent space -- the ``active subspace''. Functions of this form are known as \textit{ridge} functions \cite{pinkus2015ridge}, and can be shown to naturally arise from simple physical laws (see, e.g., the well-known Buckingham-$\pi$ theorem \cite{buckingham1914physically}), where $f$ is largely controlled by one dominant physical process that depends on several key (but latent) variables $\bm{c} = \bm{M} \bm{x}$. The challenge lies in jointly estimating and quantifying uncertainty on the embedding matrix $\bm{M}$ and the underlying function $g$ on the active subspace. Much of the existing literature considers polynomial models on $g$ (see, e.g., \cite{constantine_discovering_2015, glaws_dimension_2017, scillitoe_polynomial_2021,guan_prediction_2020}), which facilitates efficient \textit{point} estimation of $\bm{M}$ via quadratic programming. Recent works, such as \cite{seshadri_dimension_2019, bilionis2016gaussian}, explore more flexible forms of $g$ via GPs, with the goal of estimating the underlying active subspace. These are also known as \textit{multi-index} models in the literature (see \cite{hall_linear_1993,xia_mim_2008}); however, existing work on such models are again largely restricted to parametric models \cite{tyagi_mim_2012}. Another key limitation with the above existing models is that they do not quantify uncertainty on the embedding matrix $\bm{M}$ within predictive modeling. As we shall see next, this may result in wildly overconfident surrogates with poor uncertainty quantification (see \cite{yuchi2023bayesian} for similar observations).

Finally, there has been recent work (see, e.g., \cite{gilboa_scaling_2013, delbridge_randomly_2020, chen2022projection}) on identifying active low-dimensional structure via the integration of projection pursuit within a GP. Such models take the form $f(\bm{x}) = \sum_{l=1}^L g_l(\bm{m}_l^T\bm{x})$, i.e., $f$ can be represented as the sum of separate single-index models. Here, $g_l$ follows a GP, and the projection vectors $\bm{m}_l$ can either be optimized or randomly sampled. While this provides a flexible extension of the single-index GP, it suffers from a similar limitation as earlier models for our application: it does not directly elicit (and thus does not model for) the embedding structures dictated by the \textit{multi-physics} present in the simulation system. With limited data, this \textit{misspecification} of the underlying embedded structure may result in worse predictive performance compared to standard GPs, as we shall see next.




\subsection{Limitations for the QGP application}
\label{sec:limitations}

To motivate the AdMIn-GP, we investigate the above existing surrogates for our application. We first simulate the training design points using Latin hypercube designs \cite{santner2003design} of sizes $n$ from the expensive physics model, which has $d=17$ parameters. We then compare several existing models: (i) the standard GP model using the squared-exponential kernel with automatic relevance determination (ARD-GP), (ii) the fully Bayesian single-index GP model (SIM-GP; \cite{gramacy2012gaussian}), (iii) the active subspace (or dimension reduced) GP model (DR-GP; \cite{seshadri_dimension_2019, bilionis2016gaussian, snelson2012variable,bilionis2016gaussian}), which performs dimension reduction using point estimates of the active subspaces, and (iv) the projection-pursuit-based diverse projected additive GP (DPA-GP; \cite{delbridge_randomly_2020}). Model hyperparameters are fitted via maximization of the marginal likelihood, as recommended in \cite{rasmussen_occam_2000, titsias_bayesian_2010, bilionis2016gaussian}. The fitted models are then compared on its root-mean-squared-error (RMSE), the continuous ranked probability score (CRPS; \cite{gneiting2007strictly}), and the empirical coverage of its 95\% CIs over a testing set. The first measures the point prediction accuracy, the second for probabilistic predictions, and the last measures coverage performance. Further details on this set-up are provided in Section \ref{sec:app}.

\begin{figure}[!t]
\includegraphics[width=\textwidth]  {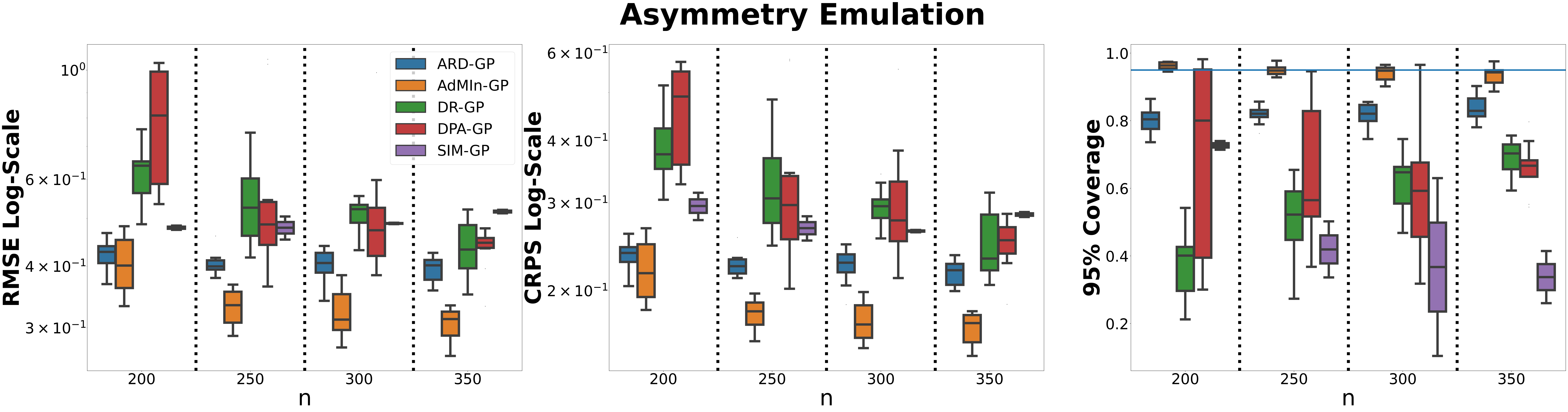}
\caption{Boxplots of different predictive metrics (left: RMSE; middle: CRPS; right: empirical coverage rate) for existing models and the proposed AdMIn-GP.}
\label{fig:appetizer}
\end{figure}

Figure \ref{fig:appetizer} shows the predictive metrics for each model with varying sample sizes $n = 200 - 500$ for our QGP application. We see that, aside from the SIM-GP, existing approaches that attempt to fit embedded low-dimensional structure surprisingly perform \textit{worse} than the standard GP, in terms of both predictions and coverage! The SIM-GP yields comparable predictions to the standard GP (in terms of RMSE and CRPS), but provides significantly worse uncertainty quantification as its coverage rate is noticeably lower than the desired 95\%. There are two plausible reasons for this. First and foremost, when the embedded low-dimensional structure (which we know to be present from prior knowledge of multi-physics) is \textit{misspecified} in the surrogate model, the fitted model may in fact yield worse predictions and coverage over standard GPs that do not leverage such structure. This highlights the need for a careful elicitation of the embedded structure from prior scientific knowledge for surrogate modeling. Second, the poor coverage of existing methods is not too surprising, since such methods largely do not account for uncertainties in the estimation of the embedding matrix $\bm{M}$. To foreshadow, Figure \ref{fig:appetizer} shows the performance of the proposed AdMIn-GP, which by modeling for the desired embedding structure guided by the multi-physics of the QGP, appears to yield significantly improved predictions with desired probabilistic coverage.

\section{The AdMIn-GP model}
\label{sec:vmegp}

\begin{figure}[!t]
\centering
\includegraphics[width=0.6\textwidth]{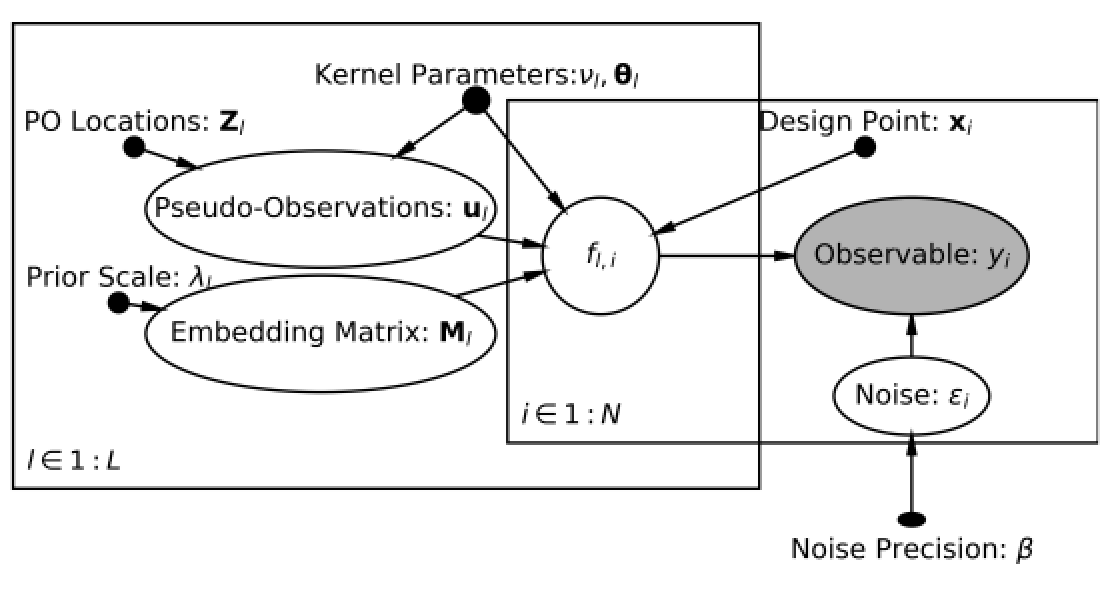}
\caption{Plate diagram visualizing the inducing-points formulation for the AdMIn-GP. White nodes represent latent (unobserved) variables, the shaded node represents observed data, black points represent input variables or model hyperparameters, and arrows indicate model dependencies. }
\label{fig:plate}
\end{figure}

We now present the proposed AdMIn-GP and justify how its modeled low-dimensional structure can capture multi-physics in the simulator. We then propose a variational inference approach using inducing points for efficient posterior inference and prediction. Figure \ref{fig:plate} visualizes our modeling framework in plate diagram form; we elaborate on each part below.

\subsection{Model specification}
\label{sec:spec}

Let us assume the same noisy observable model in \eqref{eq:obs}, with $y_1, \cdots, y_n$ the noisy simulated outputs with i.i.d. noise $\mathcal{N}(0, \beta^{-1})$ at input parameters $\bm{x}_1, \cdots, \bm{x}_n$. To capture the desired multi-physics within $f$, the AdMIn-GP adopts the following model:
\begin{equation}
    f(\bm{x}_i) = \sum_{l=1}^L f_l(\bm{x}_i) := \sum_{l=1}^L g_l(\bm{M}_l \bm{x}_i), \quad i = 1, \cdots, n.
    \label{eq:admingp}
\end{equation}
Here, for each term $l$, the embedding matrix $\bm{M}_l \in \mathbb{R}^{p \times d}$ maps the original $d$ inputs in $\bm{x}_i$ onto a lower $p$-dimensional subspace. This additive multi-index form \eqref{eq:admingp} is guided by the \textit{multi-physics} structure present in the simulator $f$, with each of the $L$ terms, namely $g_l(\bm{M}_l\bm{x}_i)$, modeling for the effect of distinct dominant physics in the simulator. As mentioned in Section \eqref{eq:gp}, each physics is likely active on a low-dimensional (but latent) embedding of the $d$ parameters; see, e.g., the Buckingham-$\pi$ theorem \citep{buckingham1914physically}. This is reflected in the fact that $g_l(\cdot)$ is active only on the lower $p$-dimensional latent variables $\bm{c}_{l} := \bm{M}_l \bm{x}$.

Of course, neither $g_l(\cdot)$ nor $\bm{M}_l$ are known in practice. For the functions $\{g_l(\cdot)\}_{l=1}^L$, we then assign independent zero-mean Gaussian process priors with kernel $k_l$, i.e.,
\begin{equation}
g_l \distas{indep.} \text{GP}\{0,k_l(\cdot,\cdot)\}, \quad l = 1, \cdots, L.
\end{equation}
This provides a flexible Bayesian nonparametric form for learning the effect of the latent variables on the response of interest. For the embedding matrices $\{\bm{M}_l\}_{l=1}^L$, we employ independent element-wise priors of the form:
\begin{equation}
(\bm{M}_l)_{jk} \distas{indep.} \textup{DExp}(0,\lambda_l), \quad j = 1, \cdots, p, \; k = 1, \cdots, d, \; l = 1, \cdots, L,
\label{eq:de}
\end{equation}
where $\textup{DExp}(0,\lambda_l)$ is the double exponential distribution with mean 0 and scale parameter $\lambda_l$. Such priors have been widely used for {shrinkage} in high-dimensional Bayesian linear regression \cite{bhattacharya2015dirichlet}. They are used here for two reasons. First, it is known from prior physical knowledge that, for each dominant physics $l$, the active latent parameters $\bm{c}_{l} := \bm{M}_l \bm{x}$ depend on a \textit{sparse} number of the original $d$ parameters. {For the QGP, such sparsity is expected from the multi-stage nature of high-energy nuclear collisions, where groups of parameters act at specific stages; it has also been observed experimentally in prior QGP studies ~\cite{sangaline2016toward,everett2021multisystem}.} Second, with limited $n$, some shrinkage of the large number of matrix parameters can prevent model overfitting. Here, double exponential priors are preferred over standard Gaussian priors since sub-exponential tails are known to overshrink parameters empirically \cite{Ray_laplace_2021, Castillo_laplace_2012}. We will discuss the estimation of $\{\lambda_l\}_{l=1}^L$ and other hyperparameters later.

For the kernels $\{k_l\}_{l=1}^L$, we found that the isotropic squared-exponential form $k_l(\bm{c}, \bm{c}') = \nu_l \exp\{ -(2\theta_l)^{-1}\|\bm{c} - \bm{c}'\|^2_2\} $ works quite well in implementation, where different variance and length-scale parameters are used for each of the $l$-th component. While such a kernel is isotropic on the latent embedded space, the induced kernel on the original $d$ parameters is anisotropic; in the simple case of $L=1$ component, $\text{Cov}\{ f(\bm{x},\bm{x}') \} = \nu_l  \exp\{ -(2\theta_l)^{-1}(\bm{x} - \bm{x}')^T \bm{M}_l^T\bm{M}_l(\bm{x} - \bm{x}')\}$, which is clearly anisotropic. With careful probabilistic estimation of the embedding matrices $\bm{M}_l$ (see next), the AdMIn-GP thus facilitates the identification of important combinations of parameters in $\bm{x}$ for predictive modeling. The choice of an isotropic kernel in this embedded space also has the key advantage of speeding up hyperparameter optimization and providing numerical stability; more on this later. Our framework can also accommodate other choices of isotropic kernels (e.g., the isotropic Mat\'ern \cite{stein1999interpolation}), with slight modifications on the following variational inference procedure.

\subsection{Inducing points and variational inference}
\label{sec:vi}

While the above model is guided by the desired multi-physics embedding structure, the many model parameters to estimate (e.g., parameters in $\bm{M}_l$) can introduce significant computational complexities. To address this, we extend the inducing points approach in \cite{snelson_sparse_2005, titsias_bayesian_2010}, which is widely used for fitting ``deep'' variants of GPs \cite{damianou_dgp_2013}, by facilitating tractable marginalization in computing the posterior predictive distribution. Similar approaches have been employed in the spatial statistics literature (see, e.g., the Gaussian predictive process models in \cite{banerjee_stat_2008}) for scaling up GP fitting for massive datasets. 

For the AdMIn-GP, the key idea is to make use of a set of $m$ \textit{inducing points} (or ``pseudo-inputs'') $\bm{z}_{1,l}, \cdots, \bm{z}_{m,l} \in \mathbb{R}^d$ for each additive component $l = 1, \cdots, L$, with corresponding latent ``pseudo-observations'' $\bm{u}_l = (f_l(\bm{z}_{1,l}), \cdots, f_l(\bm{z}_{m,l}))$, to speed up computation. These points are not necessarily a subset of the original data points $\bm{x}_1, \cdots, \bm{x}_n$, nor its responses a subset of the original responses $\bm{y}$. 
Let $\bm{f}_l := (f_l(\bm{x}_1), \cdots, f_L(\bm{x}_n))$, and define the kernel matrices $\bm{K}_{n,m}^l = {[k_l(\bm{M}_l\bm{x}_i,\bm{M}_l\bm{z}_{j,l})]_{i=1}^n}_{j=1}^m$, $\bm{K}_{m,m}^l = {[k_l(\bm{M}_l\bm{z}_{j,l},\bm{M}_l\bm{z}_{j',l})]_{j=1}^m}_{j'=1}^m$, and $\bm{K}_{n,n}^l = {[k_l(\bm{M}_l\bm{x}_i,\bm{M}_l\bm{x}_{i'})]_{i=1}^n}_{i'=1}^n$. We can then write the generative distribution on the data $\bm{y}$ as:
\begin{align}
\begin{split}
    &\bm{y} | \bm{f}_1, \dots , \bm{f}_L  \sim \mathcal{N}\left( \sum_{l=1}^L \bm{f}_l, \beta^{-1}\bm{I}_{n \times n} \right), \quad l = 1, \cdots, L,\\
    &\bm{f}_l | \bm{u}_l, \bm{M}_l \sim  \mathcal{N}\{\bm{K}_{n,m}^l(\bm{K}_{m,m}^l)^{-1}\bm{u}_l, \bm{K}_{n,n} + \bm{K}_{n,m}^l (\bm{K}_{m,m}^l)^{-1} \bm{K}_{m,n}^l\},\\
    &\bm{u}_l | \bm{M}_l \sim \mathcal{N}(0, \bm{K}_{m,m}^l),
    \end{split}
    \label{eq:indp}
\end{align}
where each entry of $\bm{M}_l$ again follows the double-exponential shrinkage priors in \eqref{eq:de}. Figure \ref{fig:plate} shows a plate representation of this inducing-points model formulation. The pseudo-observations $\bm{u} = (\bm{u}_1; \bm{u}_2;  \dots; \bm{u}_L)$ are latent variables that will be marginalized out later for posterior prediction. We present next a variational inference (VI) approach that leverages this inducing-points formulation for efficient (approximate) Bayesian predictions.

From \eqref{eq:indp}, we wish to sample from the joint posterior distribution of the function components $\bm{f} = (\bm{f}_1; \cdots; \bm{f}_L)^T$, the pseudo-observations $\bm{u}$, and the embedding matrices $\bm{M}=\{\bm{M}_1, \cdots, \bm{M}_L\}$. We leverage a similar variational bound as in \cite{titsias_bayesian_2010}, which was originally proposed for variational inference of Gaussian process latent variable models. The desired posterior is first approximated by the variational form:
\begin{align}
    p(\bm{f}, \bm{u}, \bm{M}| \bm{y}) \approx \phi(\bm{u}) \prod_{l=1}^L p(\bm{f}_l|\bm{u}_l, \bm{M}_l) \phi_l(\bm{M}_l),
    \tag{VI}
    \label{eq:approx}
\end{align}
where $p(\bm{f}_l| \bm{u}_l, \bm{M}_l)$ is the multivariate normal density in \eqref{eq:indp}. Here, $\phi(\bm{u})$ and $\phi_l(\bm{M}_l)$ are variational distributions\footnote{We use the notation $\phi(\theta)$ to denote the \textit{variational} distribution of a parameter $\theta$, to distinguish it from its generative distribution $p(\theta)$ in \eqref{eq:indp}.} that will be selected for a computationally tractable variational bound for the log-marginal likelihood of our observations, $\log p(\bm{y})$. The variational form \eqref{eq:approx} and resulting variational bound mimic those in \cite{titsias_bayesian_2010} for the Bayesian GP latent variable modeling. This form facilitates the analytical marginalization desired for optimizing the evidence lower bound (see next), while also providing flexibility for modeling dependencies between different additive components, embedding matrices and pseudo-outputs.

Variational inference then proceeds via the maximization of a lower bound on the marginal likelihood $\log p(\bm{y})$ via the variational approximation \eqref{eq:approx}. Standard VI methods for mean-field and structured approximations (see, e.g., \cite{hoffman_stochastic_2015}) are difficult to apply here due to the unstructured coupling of variables in the multivariate normal likelihood $p(\bm{f}_1, \dots, \bm{f}_L| \bm{u}, \bm{M}_1, \cdots, \bm{M}_L)$, hence more established VI approaches (see \cite{blei_variational_2017}) cannot be used. We thus need to explicitly derive below an analytic lower bound on the marginal likelihood for maximizing variational parameters in $\phi$ and GP model parameters.

With \eqref{eq:approx}, the log-marginal likelihood can be bounded via the evidence lower bound (ELBO; \cite{blei_variational_2017}):
\begin{align}
\begin{split}
    \log p(\bm{y}) &\geq \int \phi(\bm{u}) \left( \prod_{l=1}^L p(\bm{f}_l| \bm{u}_l, \bm{M}_l)\phi_l(\bm{M}_l) \right) \times\\
    & \quad \quad \quad \quad  \log\left\{ \frac{p(\bm{y}|\bm{f}_1 \dots \bm{f}_L) \prod_{l=1}^L p(\bm{f}_l| \bm{u}_l, \bm{M}_l)p(\bm{u}_l)p(\bm{M}_l)}{\phi(\bm{u})\prod_{l=1}^L p(\bm{f}_l| \bm{u}_l, \bm{M}_l)\phi(\bm{M}_l)} \right\} d\bm{f} d\bm{M} d\bm{u}.
    \label{eq:elbo1}
\end{split}
\end{align}
We can then marginalize out the latent function components $\bm{f}$, yielding:
\begin{align}
\begin{split}
    &\int \phi(\bm{u}) \left( \mathbb{E}_{\phi(\bm{M})}\left[\log \mathcal{N}\left(\bm{y}; \sum_{l=1}^L \boldsymbol \alpha_l , \beta^{-1} \bm{I}_{n \times n}\right) \right] + \log\frac{p(\bm{u})}{\phi(\bm{u})} \right) d\bm{u} \\
    & \quad \quad - \sum_{l=1}^L \frac{\beta}{2}\left(\nu_l n - \text{tr}\left[(\bm{K}_{m,m}^l)^{-1} \mathbb{E}_{\phi(\bm{M})}(\bm{K}_{m,n}^l\bm{K}_{n,m}^l) \right] \right) - \sum_{l=1}^L \text{KL}\{\phi(\bm{M}_l) || p(\bm{M}_l)\},
    \end{split}
    \label{eq:elbo2}
\end{align}
where $\boldsymbol \alpha_l = \bm{K}_{n,m}^l (\bm{K}_{m,m}^l)^{-1} \bm{u}_l$, $p(\cdot)$ denotes the generative distribution of a parameter from \eqref{eq:indp}, and $\text{KL}$ denotes the Kullback-Liebler divergence. Examining the first integral, note that if the variational distribution on $\bm{u}$ is chosen as:
\begin{equation}
\phi(\bm{u}) \propto \exp\left\{ \mathbb{E}_{\phi(\bm{M})}\left(\log \mathcal{N}\left(\bm{y};\sum_{l=1}^L \boldsymbol \alpha_l , \beta^{-1} \bm{I}_{n \times n}\right)\right) \right\}p(\bm{u})
\label{eq:phiu}
\end{equation}
then the bound in \eqref{eq:elbo2} reduces to:
\begin{align}
\small
\begin{split}
\log p(\bm{y}) &\geq  \int p(\bm{u}) \mathbb{E}_{\phi(\bm{M})}\left[ \log\mathcal{N}\left(\bm{y};\sum_{l=1}^L \boldsymbol \alpha_l , \beta^{-1} \bm{I}_{n \times n}\right)\right] d\bm{u}\\
    & \quad - \sum_{l=1}^L \frac{\beta}{2}\left(\nu_l n - \text{tr}\left[(\bm{K}_{m,m}^l)^{-1} \mathbb{E}_{\phi(\bm{M})}(\bm{K}_{m,n}^l\bm{K}_{n,m}^l) \right] \right) - \sum_{l=1}^L \text{KL}\{\phi(\bm{M}_l) || p(\bm{M}_l)\}.
\end{split}
\label{eq:elbo3}
\end{align}

This lower bound can now be computed via simple integration of the pseudo-observations $\bm{u}$ and the expected matrices $\boldsymbol{\Psi}_1^l = \mathbb{E}_{\phi(\bm{M})}(\bm{K}_{n,m}^l), \boldsymbol{\Psi}_2^l = \mathbb{E}_{\phi(\bm{M})}(\bm{K}_{m,n}^l \bm{K}_{n,m}^l)$.  We first choose the variational distribution on $\bm{M}$ as the matrix normal distribution \cite{chikuse2003statistics}, i.e.,
\begin{equation}
    \phi_l(\bm{M}_l) \sim \mathcal{MN}(\bm{A}_l, \bm{H}_l, \bm{U}_l), \quad l = 1, \cdots, L,
    \label{eq:phim}
\end{equation}
where $\bm{A}_l$, $\bm{H}_l$ and $\bm{U}_l$ are the mean, row-wise covariance and column-wise covariance matrices. With this, and setting $k_l$ as the isotropic squared-exponential kernel, each element of $\boldsymbol{\Psi}_1^l$ and $\boldsymbol{\Psi}_2^l$ can then be computed in closed form. The derivations are, however, quite tedious and are deferred to Appendix \ref{appendix:integral}. Integrating these closed-form expressions for $\boldsymbol{\Psi}_1^l$ and $\boldsymbol{\Psi}_2^l$ into \eqref{eq:elbo3}, we arrive at the final variational bound for the log-marginal likelihood:
\begin{align}
\begin{split}
    \log p(\bm{y}) &\geq  \log\left( \frac{\beta^{\frac{n}{2}}\prod_{l=1}^L \det(\bm{K}_{m,m}^l)^{\frac{1}{2}}}{(2\pi)^{\frac{n}{2}}\det(\beta \bm{D} + \bm{P})^{\frac{1}{2}}} \exp\left\{\frac{\beta^2}{2}\bar{\bm{y}}^T \bm{W}^{-1}\bar{\bm{y}}\right\}\exp\left\{-\frac{1}{2}\beta\bm{y}^T\bm{y} \right\} \right) \\
    & \quad \quad - \sum_{l = 1}^L \text{KL}\{\phi_l(\bm{M}_l)\} ||p(\bm{M}_l)) - \sum_{l=1}^L \text{tr}(\bm{V}_l),
    \end{split}
    \label{eq:elbo4}
\end{align}
where $\bm{V}_l = ({\beta}/{2}) (\nu_l n - \text{tr}\{(\bm{K}_{m,m}^l)^{-1} \boldsymbol{\Psi}_2^l\} )$, $\bar{\bm{y}} = ((\boldsymbol{\Psi}^1_1)^T \bm{y}, (\boldsymbol{\Psi}^2_1)^T \bm{y} \dots (\boldsymbol{\Psi}^L_1)^T \bm{y})^T$, and $\bm{W} =  \beta \bm{P} + \bm{D}$. Here, $\bm{D} \in \mathbb{R}^{mL \times mL}$ is block diagonal with $\bm{D}_{l,l} = \bm{K}_{m,m}^l$, and $\bm{P} \in \mathbb{R}^{mL \times mL}$ is a block matrix with block diagonals $\bm{P}_{l,l} = \boldsymbol{\Psi}^l_2$ and block off-diagonals $\bm{P}_{s,t} = (\boldsymbol{\Psi}^s_1)^T(\boldsymbol{\Psi}^t_1)$. Detailed derivation of this bound can be found in Appendix \ref{appendix:ELBO}

Finally, with \eqref{eq:elbo4}, we then optimize this bound by maximizing with respect to the variational parameters $\Theta_{\phi} :=\{(\bm{A}_l,\bm{H}_l,\bm{U}_l)\}_{l=1}^L$, pseudo-inputs $\mathcal{Z} :={\{\bm{z}_{j,l}\}_{j=1}^M}_{l=1}^L$, and AdMIn-GP model parameters $\Theta_{\rm GP}$, namely, its scale parameter $\nu_l$, length-scale parameter $\theta_l$ and shrinkage parameter $\lambda_l$ for each additive component $l = 1, \cdots, L$. Details on this optimization are provided in Section \ref{sec:opt}. The variational distribution \eqref{eq:approx} with these optimized parameters then serves as an approximation for the complex posterior $p(\cdot|\bm{y})$, which facilitates efficient posterior predictions (see Section \ref{sec:post}). Since the derived bound \eqref{eq:elbo4} lower bounds the log-marginal likelihood, the maximization of this bound with respect to model parameters can be viewed as an empirical Bayes estimation \cite{carlin1997bayes} of such parameters.



One bottleneck that arises is that, as each evaluation of the variational bound \eqref{eq:elbo4} requires $\mathcal{O}(L^3(nm^2d^3p^3 + m^3))$ operations, the optimization over the full matrix normal variational family \eqref{eq:phim} can be computationally quite costly. One solution is to restrict the variational family \eqref{eq:phim} to the simpler form $\phi_l(\bm{M}_l) \sim \mathcal{MN}(\bm{A}_l, \bm{I}_{p \times p}, \bm{I}_{d \times d}\bm{v}_l)$, so that only the mean matrices $\bm{A}_l$ and column-wise variances $\bm{v}_l$ need to be optimized for variational inference. In addition to reducing the number of variational parameters for optimization, this restriction facilitates the use of rank-one matrix updates to achieve improved computational complexity per evaluation of the bound \eqref{eq:elbo4}, namely $\mathcal{O}(L^3(nm^2pd + m^3))$. Technical details of this simplification are provided in Appendix \ref{appendix:integral}. In our later experiments, this simplified variational form does not appear to compromise much uncertainty quantification performance, particularly in comparison to existing models.



\subsection{Posterior predictions}
\label{sec:post}
Finally, we can make use of the variational form \eqref{eq:approx} (with variational parameters optimized via the ELBO) to approximate the desired posterior predictive distribution $[f(\bm{x}_{\rm new}) | \bm{y}]$, where $\bm{x}_{\rm new}$ is a new input point. Note that our choice of variational distributions in \eqref{eq:phiu} and \eqref{eq:phim} induces the marginal distribution on pseudo-observations:
\begin{equation}
\phi(\bm{u})  \sim \mathcal{N}(\bm{D} \bm{W}^{-1} \bar{\bm{y}}, \bm{D} \bm{W}^{-1}\bm{D}).
\label{eq:indmarg}
\end{equation}
Integrating this within \eqref{eq:approx}, one can then show that $[f(\bm{x}_{\rm new}) | \bm{y}, \bm{M}]$, the posterior predictive distribution conditional on the embedding matrices $\bm{M}$, can be approximated as:
\begin{equation}
        \mathcal{N}\Big( \bm{K}_{{\bm{x}_{\rm new}}, \bm{M}} \bm{D}^{-1} \boldsymbol \mu, \bm{K}_{\bm{x}_{\rm new}, \bm{x}_{\rm new} } - \bm{K}_{\bm{x}_{\rm new}, \bm{M}}\bm{D}^{-1}\bm{K}_{\bm{M}, \bm{x}_{new}} + \bm{K}_{\bm{x}_{\rm new}, \bm{M}}\bm{D}^{-1} \bm{G} \bm{D}^{-1} \bm{K}_{\bm{M}, \bm{x}_{\rm new}} \Big).
        \label{eq:predcond}
\end{equation}
Here, $\boldsymbol \mu = \bm{D}\bm{W}^{-1} \bar{\bm{y}}$, $\bm{G} = \bm{D}\bm{W}^{-1}\bm{D}$, and $\bm{K}_{\bm{x}_{\rm new}, \bm{M}}$ is a vector of length $mL$ such that its subvector (from indices $l$ to $m(l+1)-1$) is the cross-covariance vector between the new point $\bm{x}_{\rm new}$ and the $m$ inducing points for additive component $l$ given embedding matrices $\bm{M}$ (see \eqref{eq:indp}). With this, we can then approximate the desired posterior distribution by marginalizing over $\phi(\bm{M}) = \prod_{l=1}^L \phi(\bm{M}_l)$, the variational distribution on $\bm{M}$ in \eqref{eq:phim}:
\begin{equation}
[f(\bm{x}_{\rm new}) | \bm{y}] = \int [f(\bm{x}_{\rm new}) | \bm{M}, \bm{y}] \; [\bm{M}|\bm{y}] \; d\bm{M} \approx \int \phi(f(\bm{x}_{\rm new})|\bm{M})  \phi(\bm{M}) \; d\bm{M},
\label{eq;predmarg}
\end{equation}
where $\phi(f(\bm{x}_{\rm new})|\bm{M})$ is the multivariate normal distribution in \eqref{eq:predcond}. One can thus sample from this approximate posterior predictive distribution, by first sampling the embedding matrices $\bm{M}$ from $\phi(\bm{M})$ in \eqref{eq:phim}, then sampling the prediction $f(\bm{x}_{\rm new})$ from $\phi(f(\bm{x}_{\rm new})|\bm{M})$ in \eqref{eq:predcond}. Both sampling steps are multivariate normal, and thus can be performed efficiently after variational parameters are optimized via ELBO.


The above modeling approach for the AdMIn-GP highlights several advantages over the state-of-the-art, particularly in addressing the limitations from Section \ref{sec:limitations} for our application. First, the proposed model captures the desired multi-physics structure in the computer simulator, via an additive multi-index model with each additive component representing distinct physics active on a low-dimensional manifold. With a careful selection of the additive components and manifold ranks, the fitted model can potentially extract \textit{interpretable} physical features (see Section \ref{sec:app}) that enable improved predictions with limited data. Second, our approach quantifies \textit{uncertainty} on the underlying embedding matrices via a variational inference approach with shrinkage priors on $\bm{M}$. This addresses a key weakness of the existing models in Section \ref{sec:gp}, which largely do not account for uncertainty on $\bm{M}$ and thus can yield poor coverage for our application (Section \ref{sec:limitations}). Finally, the proposed approach enables \textit{efficient} predictions via a carefully-constructed variational approximation of the posterior predictive distribution; we will see later in numerical experiments that this facilitates quick and accurate surrogate modeling with rich uncertainty quantification.


\section{Model Implementation}
\label{sec:impl}

We now discuss important implementation details of the AdMIn-GP that will be employed in numerical experiments. We first present an approach for optimizing parameters in the variational lower bound \eqref{eq:elbo4}, then discuss efficient model selection approaches for fitting the number of additive components $L$ and the rank of embedding matrices $p$.


\subsection{Parameter optimization}
\label{sec:opt}

Recall that the optimization of the variational bound \eqref{eq:elbo4} facilitates the estimation of variational parameters $\Theta_{\phi}:=\{(\bm{A}_l,\bm{v}_l)\}_{l=1}^L$, pseudo-inputs $\mathcal{Z} :={\{\bm{z}_{j,l}\}_{j=1}^M}_{l=1}^L$, and the AdMIn-GP parameters $\Theta_{\rm GP} := \{(\nu_l,\theta_l,\lambda_l)\}_{l=1}^L$. Letting $\text{VLB}(\Theta_{\phi},\mathcal{Z},\Theta_{\rm GP})$ denote the variational lower bound in \eqref{eq:elbo4}, this optimization can be stated as:
\begin{equation}
(\hat{\Theta}_{\phi}, \hat{\mathcal{Z}}, \hat{\Theta}_{\rm GP}) = \argmax_{\Theta_{\phi}, \mathcal{Z}, \Theta_{\rm GP}} \text{VLB}(\Theta_{\phi},\mathcal{Z},\Theta_{\rm GP}).
\label{eq:opt} 
\end{equation}
The optimized parameters $(\hat{\Theta}_{\phi}, \hat{\mathcal{Z}}, \hat{\Theta}_{\rm GP})$ are then plugged into \eqref{eq:approx} for posterior approximation. In this formulation, both the inputs and response should be standardized to zero mean and unit variance; we found this to be essential for stable optimization. Due to the sheer number of parameters to optimize, we further employed a common set of inducing points $\{\bm{z}_j\}_{j=1}^m$ over all additive components, with common shrinkage and scale parameters $\lambda$ and $\nu$. In our later experiments, such simplifications appeared to yield significant computational speed-up without noticeable compromise in predictive and UQ performance.

The high-dimensional optimization problem \eqref{eq:opt}, however, can be highly challenging. While analytic gradients can be derived from \eqref{eq:approx}, the sheer number of parameters makes such gradients burdensome to compute and carry around for optimization. We thus leverage recent tools on automatic differentiation \cite{paszke2017automatic}, which is widely used for efficient training of complex machine learning models. The key idea is to exploit the computation of $\text{VLB}(\cdot)$ as a sequence of elementary operations, to provide accurate gradient estimates with a single forward evaluation of the objective. In our implementation, we made use of the automatic differentiation capabilities in PyTorch \cite{paszke2017automatic}. The ADAM optimizer \cite{kingma_adam_2014}, a popular optimizer in machine learning, is then used for solving \eqref{eq:opt}; we found that a learning rate of 0.005 with 4000 optimization steps yielded acceptable performance, with a smaller learning rate of 0.0025 if divergent behavior arises. We further made use of GPU computing architecture to speed up matrix multiplication steps involved in evaluating the objective function $\text{VLB}(\cdot)$.




A careful initialization of parameters is also important for successful optimization. For the variational parameters $\Theta_{\phi}$, we found that a random initialization $\bm{M}_l \sim \mathcal{N}(\boldsymbol{0}, \bm{I}_{p \times p}, \bm{I}_{d \times d})$ is quite effective for the embedding matrices. The common set of inducing points is initialized as $\bm{z}_m \distas{i.i.d.} \mathcal{N}(\boldsymbol{0}, \bm{I}_{m \times m})$, with column-wise variances initialized at $\bm{v}_l = 0.1$. For the GP model parameters $\Theta_{\rm GP}$, we initialized the common shrinkage parameter $\lambda$ at a relatively small value 0.01, and the common scale parameter $\nu$ uniformly between $[1,2]$.


\subsection{Model selection}
\label{sec:model_selection}

Another important step is model selection, namely, the selection of the number of additive components, $L$, and the rank of the embedded subspaces, $p$. A standard approach would be to estimate such parameters via cross-validation \cite{hastie2009elements}, but with the complexity of our model, this can be prohibitive in terms of computational cost.  Cross-validation may further induce larger biases in the current limited data regime. We instead perform the model selection of $L$ and $p$ via the maximization of the variational bound \eqref{eq:elbo4}. More formally, letting $(\hat{\Theta}_{\phi}(L,p), \hat{\mathcal{Z}}(L,p), \hat{\Theta}_{\rm GP}(L,p))$ denote the optimized parameters from \eqref{eq:opt} given fixed $L$ and $d$, we select the optimal $L^*$ and $p^*$ via:
\begin{equation}
(L^*,p^*) = \argmax_{L,p} \text{VLB}(\hat{\Theta}_{\phi}(L,p), \hat{\mathcal{Z}}(L,p), \hat{\Theta}_{\rm GP}(L,p)).
\label{eq:modsel} 
\end{equation}
As $\text{VLB}$ lower bounds the log-marginal likelihood of the data $\bm{y}$ (see \eqref{eq:elbo4}), such a procedure can be viewed as an approximate empirical Bayes model selection approach under uniform (flat) priors over feasible combinations of $(L,p)$ \cite{hoffman_empirical_2013}. In general, we find that selection of embedding dimension $p$ to be less important than the number of additive components $L$; the shrinkage priors on $\bm{M}_l$, coupled with known sparsity of the multi-physics phenomena (see Section \ref{sec:spec}), results in similar empirical performance over different choices of $p$.


 
In implementation, we found that to ensure numerical stability, the embedding rank $p$ should be set above a lower limit of 4 for model selection. An upper limit on $p$ does not appear necessary though, since the employed shrinkage priors on $\bm{M}$ facilitate stable model fits for large $d$. For the number of inducing points $m$, one should choose as many points as possible, but this can be quite limited due to computation. For our later experiments, we made use of $m=125$ inducing points, which seemed to provide adequate performance.


\section{Numerical Experiments}
\label{sec:exp}
We now investigate the predictive and UQ performance of the proposed AdMIn-GP. We first discuss the simulation set-up, including evaluation metrics and compared models. We then explore the effectiveness of the AdMIn-GP in a suite of simulation experiments.

 \subsection{Simulation set-up}
 \label{sec:evaluation}

We compare with the same four models tested in Section \ref{sec:limitations}:
\begin{itemize}
\item The standard GP model with the squared-exponential kernel and automatic relevance determination (ARD-GP; see \cite{gramacy_surrogates_2020}), implemented in the \texttt{GPy} package \cite{gpy2014} in Python,
\item The fully Bayesian single-index model (SIM-GP; \cite{gramacy2012gaussian}), implemented in the \textsc{R} package \texttt{tgp} \cite{gramacy2007tgp},
\item The dimension-reduced (or active subspace) GP (DR-GP; \cite{seshadri_dimension_2019, snelson2012variable, bilionis2016gaussian}), with point estimates for the active subspace, implemented in the Python package \texttt{GPyTorch} \cite{gardner2018gpytorch}.


\item The projection-pursuit-based diverse projected additive GP (DPA-GP; \cite{delbridge_randomly_2020}).
\end{itemize}
The first is a standard baseline for surrogate modeling, and the latter surrogate models integrate some form of dimension reduction. The AdMIn-GP is fitted following the variational inference framework in Section \ref{sec:vmegp} and the optimization approach in Section \ref{sec:impl}. 

We then generate the training design points $\{\bm{x}_i\}_{i=1}^n$ from a Latin hypercube design \cite{mckay2000comparison}, with varying sample sizes of $n = 300$,  $400$ and $500$. The number of input parameters $d$ is set to be 20 to reflect our application (this will be increased in the next subsection). This procedure is replicated 10 times to provide a quantification of simulation variability. The embedded manifold dimension $p$ for DR-GP and DPA-GP are selected in an empirical Bayes fashion to maximize the marginal likelihood (or ELBO). Details for selection of $(L,p)$ for AdMIn-GP can be found in Section \ref{sec:mvs}.

We then evaluate the performance of these models on several metrics. The first is the standard root-mean-squared-error (RMSE) over a uniformly-sampled test set of size 1000, which measures point prediction accuracy. The second is the widely-used continuous ranked probability score (CRPS; \cite{gneiting2007strictly}), which measures accuracy of probabilistic predictions. For both measures, smaller values indicate better predictive performance. Here, the CRPS is computed from 500 samples drawn from the posterior predictive distribution of each method \cite{hersbach_crps_2000}. The third metric measures the empirical coverage rate of the 95\% posterior predictive intervals over the same test set, and coverage close to the nominal 95\% rate is desired.

\subsection{Prediction and coverage}

The base simulation case considers the following synthetic function on $f$, composed of $L=3$ additive components that each depend on their separate $p=2$-dimensional subspaces.
\begin{align}
\begin{split}
    f(\bm{x}) &= 0.4f_1(\bm{M}_1\bm{x}) + 0.3f_2(\bm{M}_2\bm{x}) + 0.3f_3(\bm{M}_3\bm{x}),\\
    f_1(\bm{c}) &= \sin(c_1) + \cos(0.5c_2), \quad f_2(\bm{c}) = \cos(0.5c_1c_2), \quad
    f_3(\bm{c}) = c_2 \cos(0.75 c_1).
    \end{split}
\end{align}
We will use this to mimic the presence of dominant multi-physics in the response surface, where each additive component captures a separate physics that depends on a low-dimensional embedding (see \cite{constantine_active_2014}). To capture this, the embedding matrices $\{\bm{M}_l\}_{l=1}^3$ are generated with $(i,j)$-th entry sampled independently from $\mathcal{N}(3, 1)$ if $l \lfloor {d}/{3} \rfloor \leq j < (l+1) \lfloor {d}/{3} \rfloor$ and $\mathcal{N}(0, 1)$ otherwise, such that each additive component is dominantly influenced by different subsets of parameters. Its rows are then normalized to unit variance. We set the noise standard deviation $\sigma$ to be 15\% of the standard deviation for $f(\bm{u})$, where $\bm{u} \sim \mathcal{N}(\boldsymbol{0},\bm{I})$.

 

\begin{figure}[!t]
         \includegraphics[width=\textwidth]{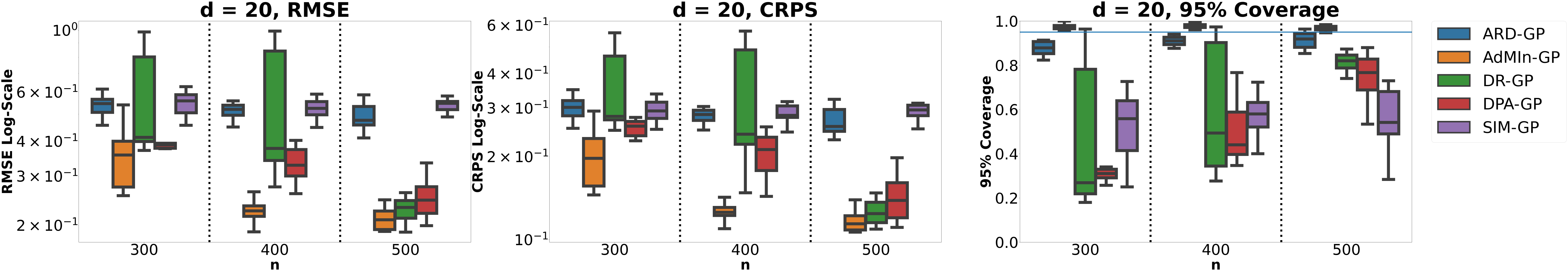}
         \caption{Boxplots of the predictive metrics for the base simulation in $d=20$ dimensions over various sample sizes $n$.}
    \label{fig:mix_pred}
\end{figure}

Figure \ref{fig:mix_pred} shows the prediction metrics for the compared models in this base simulation study. We see that the proposed AdMIn-GP yields the best prediction metrics of all models: it provides notably lower RMSE and lower CRPS, thus indicating more accurate point and probabilistic predictions. Similarly, the empirical coverage rates for the AdMIn-GP is much closer to the desired 95\% rate, whereas existing models yield markedly lower coverage; this is not too surprising, since much of these methods do not account for estimation uncertainty of the embedded manifold structure. This confirms a previous observation: when the underlying low-dimensional structure is misspecified (i.e., in DR-GP and DPA-GP), the resulting model may yield noticeably poorer predictions and coverage over the standard GP, which does not leverage such structure! When this embedding is carefully elicited, the resulting model (i.e., AdMIn-GP) can harness such structure to provide \textit{accurate} predictions and \textit{reliable} uncertainties with \textit{limited} training data -- a key challenge in our application.

\begin{figure}[!t]
         \includegraphics[width=\textwidth]{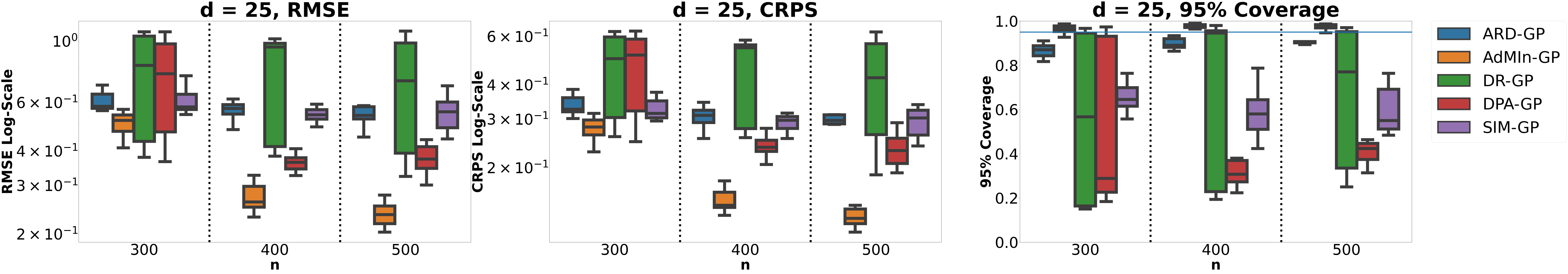}\\
         
         \includegraphics[width=\textwidth]{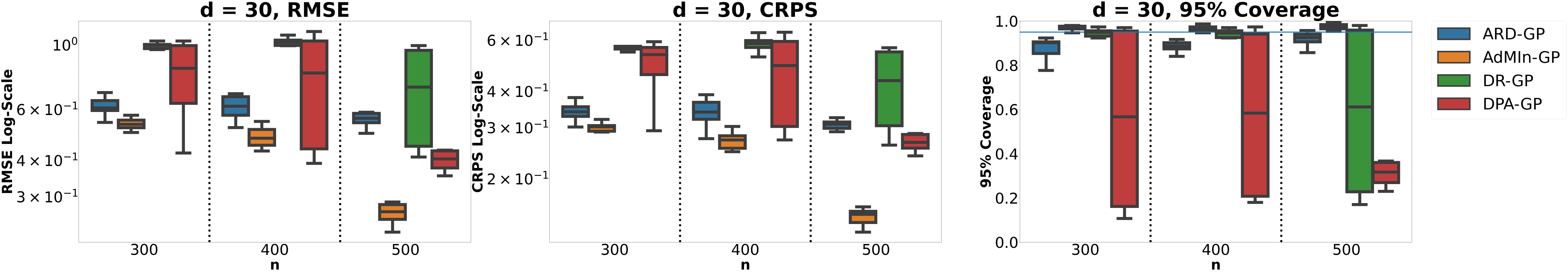}
        \caption{Boxplots of the predictive metrics for the simulations in $d=25$ and $d=30$ dimensions over various sample sizes $n$.}
        \label{fig:mix_pred2}
\end{figure}

We now investigate the effect of increasing the number of parameters $d$. Figure \ref{fig:mix_pred2} shows the predictive metrics for the compared models for two larger choices of $d = 25$ and $d = 30$. We see that similar results hold: the AdMIn-GP provide markedly more accurate point and probabilistic predictions, with empirical coverage rates closer to the desired 95\%. An interesting observation is that, as $d$ increases, the models with misspecified low-dimensional structure (DR-GP and DPA-GP) yield increasingly poorer predictive performance over the standard GP (ARD-GP). This again suggests that, for complex systems with \textit{many} input parameters, the careful elicitation and integration of known low-dimensional structures become increasingly crucial for effective surrogate modeling with limited training data. The AdMIn-GP provides a means for facilitating this integration for cost-efficient emulation of \textit{multi-physics} systems, as we show later in our motivating high-energy physics application.


\subsection{Model and variable selection}
\label{sec:mvs}

As the modeled embedded structure in the AdMIn-GP is guided by the underlying multi-physics, the estimation of certain model parameters (e.g., the number of additive components $L$ and the embedding matrices $\{\bm{M}_l\}_{l=1}^L$) may in turn shed light on the underlying dominant physics within the simulator. For example, in our QGP application, the selection of $L$ may inform the degree of physics heterogeneity in the heavy-ion collision system, and the estimation of $\{\bm{M}_l\}_{l=1}^L$ can then help identify the nature of specific dominant physics. With careful discussions with domain scientists, this may serve as a useful tool for guiding scientific discovery, as we show later in Section \ref{sec:app}. 

\begin{figure}[!t]
\includegraphics[width=\textwidth]{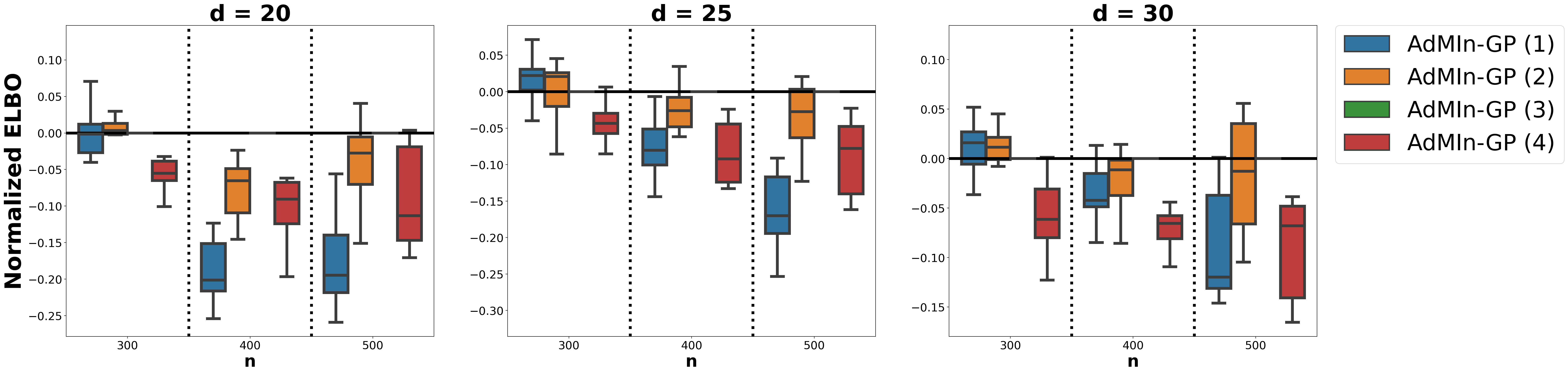}
\caption{Boxplots of the ``normalized'' ELBO bound \eqref{eq:elbo4} for the AdMIn-GP with $L$ additive components, over various sample sizes $n$. Negative values suggest that the ELBO model selection procedure prefers the $L=3$ component model.}
\label{fig:mix_choose}
\end{figure}

Consider first the model selection performance for the number of additive components $L$ in the earlier simulations. Figure \ref{fig:mix_choose} shows the boxplots of the ``normalized'' ELBO bound over different choices of $L$, in dimensions $d=20, 25$ and 30. For each of $L = 1, 2, 3$ and $4$, we set the embedding matrix dimension $p$ for each component to be $p = 12, 6, 4$ and 4, respectively, so that each model has access to approximately the same number of features, namely 12.\footnote{For the $L=4$ case, we used $p=4$ since the setting of $p=3$ yielded numerical instability issues.} This ``normalized'' ELBO bound is the variational bound \eqref{eq:elbo4} with a fixed number of components $L$, subtracted from the same variational bound with $L=3$ (the true number of components). We thus select the model with the largest ELBO, which lower bounds the marginal likelihood. For small sample sizes ($n = 300$), while the procedure tends to select the correct model with $L=3$ components, it may err by selecting a smaller number of components, particularly for larger $d$. This is not too surprising, since there are simply insufficient data to justify the more complex model. As more data are collected, we see that the correct model with $L=3$ is consistently selected, which is as desired.

Consider next the estimation of the embedding matrices $\{\bm{M}_l\}_{l=1}^L$. An important question is the \textit{identifiability} of such matrices from the data $\bm{y}$. Indeed, since the GP length-scale parameters $\{\theta_l\}_{l=1}^L$ are estimated from data, any scaling of the embedding matrix $\bm{M}_l$ by a constant would be unidentifiable, as it can be offset by a corresponding scaling of the length-scale parameter. Fortunately, for the aforementioned goal of extracting dominant physics, we require only the \textit{selection} of non-zero entries in $\{\bm{M}_l\}_{l=1}^L$ to pinpoint 
important variables in $\bm{x}$ contributing to different multi-physics (see Section \ref{sec:extract}). While we include no consistency results of this flavor for the AdMIn-GP (which are beyond the scope of the paper), there is a rich literature on the theoretical consistency and empirical performance of variable selection using GPs (see \cite{linkletter2006variable,paananen2019variable,jiang2021variable}), which suggests that the proposed model might be useful for this physics extraction goal. We investigate this empirically below.

\begin{figure}[!t]
    \centering
    \includegraphics[width = 0.8\textwidth]{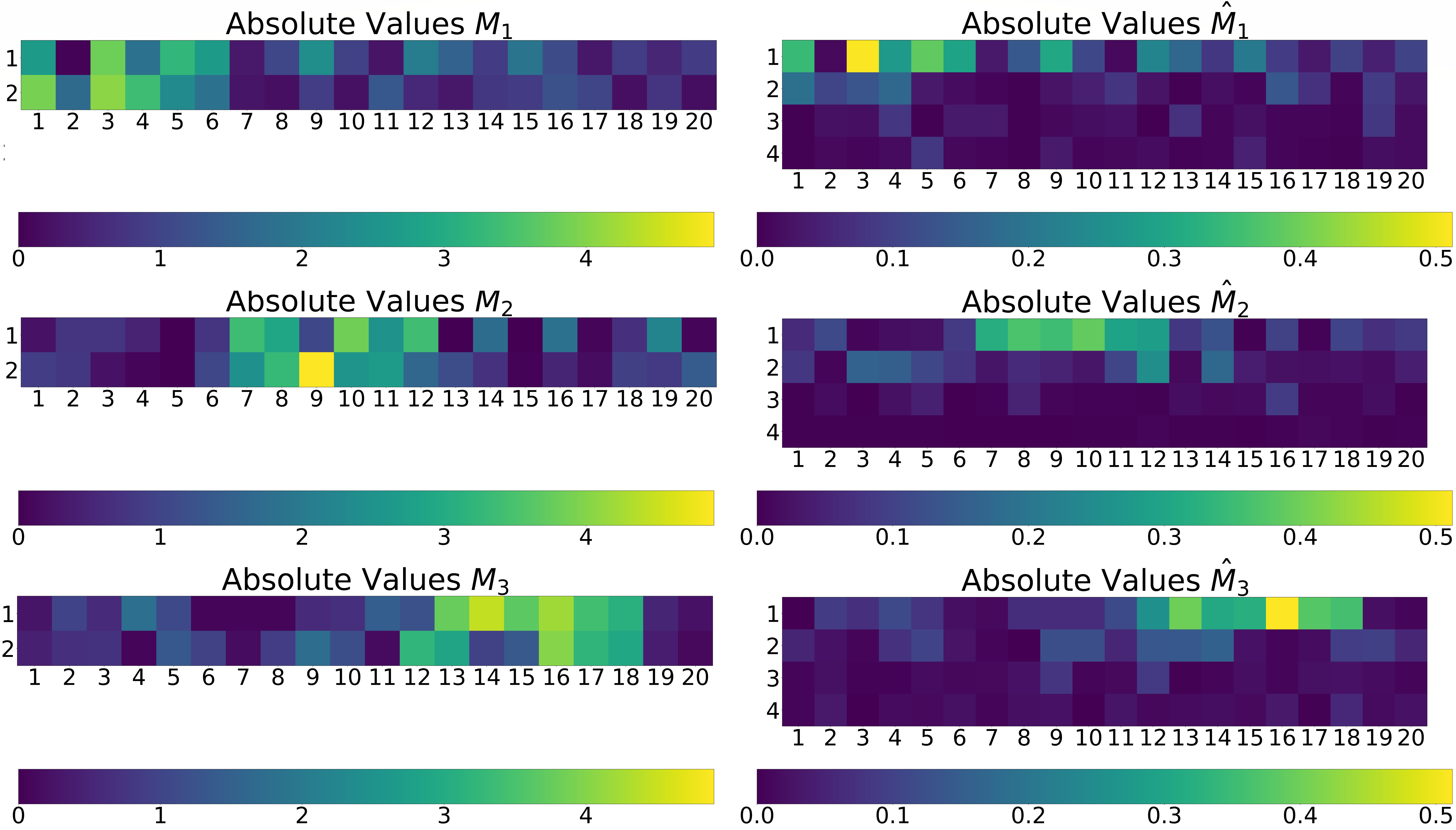}
    \caption{ (Left): Absolute values of the true (simulated) matrices $\bm{M}_1$, $\bm{M}_2$ and $\bm{M}_3$ in $d = 20$ dimensions. (Right): Absolute values of the estimated matrices $\hat{\bm{M}}_1$, $\hat{\bm{M}}_2$ and $\hat{\bm{M}}_3$ via posterior means using the proposed variational inference approach. }    
\label{fig:Mex}
\end{figure}

We now inspect how well our variational approach can identify important variables within the three embedding matrices $\bm{M}_1$, $\bm{M}_2$ and $\bm{M}_3$ in the earlier simulations. Figure \ref{fig:Mex} (left) shows the absolute values of the true matrices for the experiment in $d=20$ dimensions, where each matrix is influenced by a sparse number of parameters (as expected in multi-physics applications; see Section \ref{sec:spec}). Figure \ref{fig:Mex} (right) shows the absolute values of the estimated matrices $\hat{\bm{M}}_1$, $\hat{\bm{M}}_2$ and $\hat{\bm{M}}_3$ (via posterior means) using the proposed variational inference approach. We see that, with selected manifold rank $p=4$, each of the estimated matrices roughly identified the correct block of active parameters. For $\hat{\bm{M}}_1$, we see that the AdMIn-GP correctly identifies the first block of six variables as active, with other variables largely set as inert; similar conclusions hold for $\hat{\bm{M}}_2$ and $\hat{\bm{M}}_3$. This identification of important variables in the embeddings can help guide the discovery of interpretable dominant physics embedded in the system, as we show next. 




\section{Surrogate Modeling of the Quark-Gluon Plasma}
\label{sec:app}
We now return to our motivating application for the QGP, where the goal is to train a probabilistic emulator for simulated heavy-ion collision observables. Here, we investigate collisions of gold nuclei at near-light speeds (i.e., ultrarelativistic Au+Au collisions), which have been studied experimentally at the Relativistic Heavy Ion Collider in Brookhaven National Laboratory. 
The observables (responses) of interest are the so-called anisotropic flow coefficients $v_k$'s, which quantify \textit{momentum anisotropy} in the number of particles emitted in the transverse plane. Letting $\phi$ be the angle in the transverse plane, the particle number $dN/d\phi$ is expanded in a Fourier series and the $v_k$'s are defined as:
\begin{equation}
    \frac{d N}{d \phi}=\left\langle \frac{d N}{d \phi} \right\rangle \left[ 1 + 2 \sum_{k=1}^{\infty} v_k \cos\left(k\left(\phi-\Psi_k\right)\right) \right]
\end{equation}
where $\langle d N/d \phi \rangle$ is the average number of particles produced in a given collision, $v_k$ are its Fourier coefficients and $\Psi_k$ are the event plane angles~\cite{Heinz:2013th}. Measurements of $v_k$ are averaged over thousands of collisions, and calculations are averaged in a similar manner.

\begin{figure}[!t]
    \centering
    \includegraphics[width = .5\textwidth]{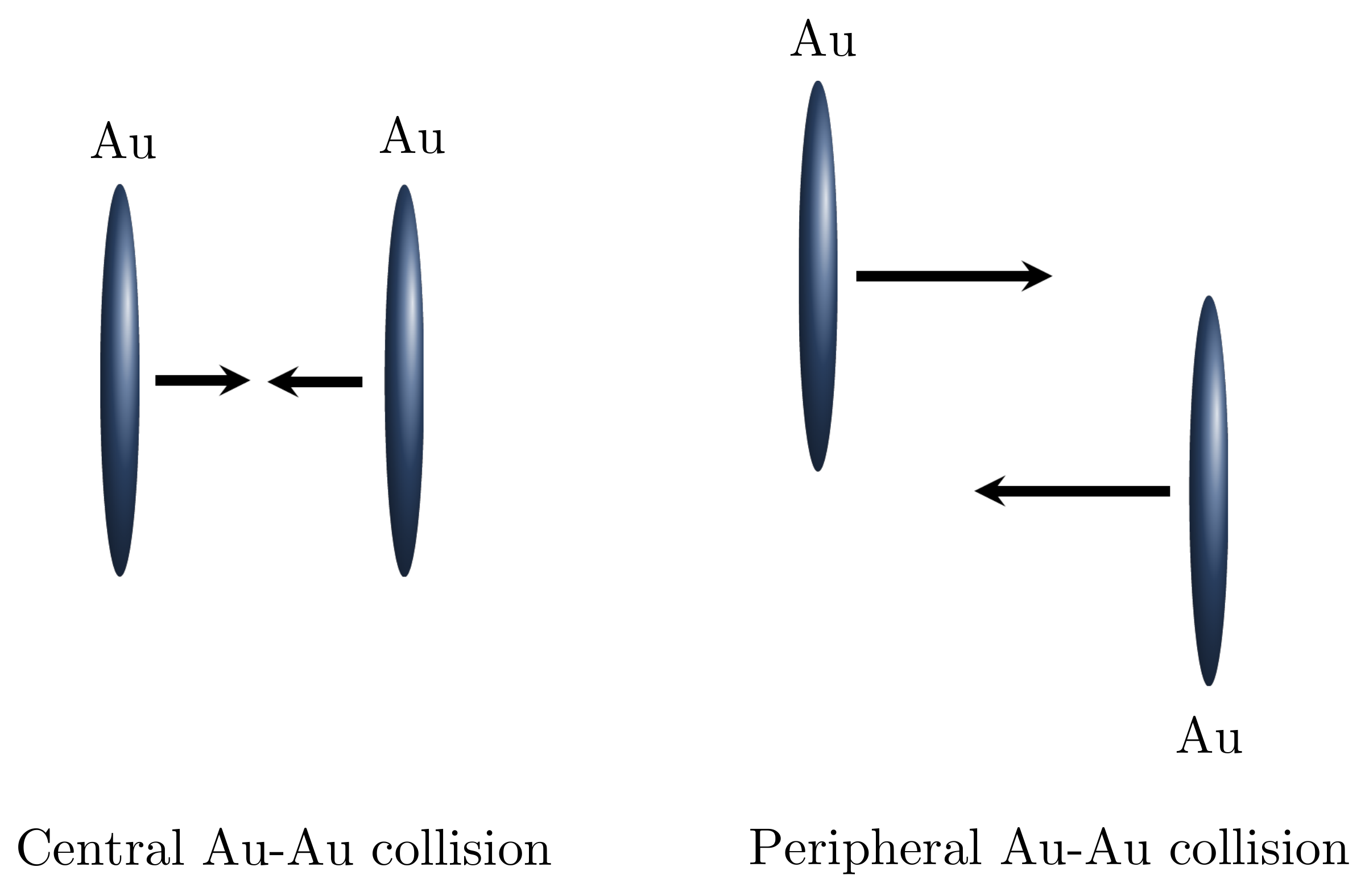}
    \caption{Visualizing a central nucleus-nucleus collision (left) where the nuclei undergo near head-on collisions, resulting in high centrality, and a {peripheral} collision (right) where the nuclei collide off-center, resulting in low centrality.}    
\label{fig:centrality}
\end{figure}

An important discovery in heavy-ion collisions is the correlation between the initial {spatial} geometry of colliding nuclei and the final {momentum} asymmetry of the number of particles at the end of the collisions (see \cite{Luzum:2013yya} for a review). As such, we use the second and third Fourier coefficients $v_2$ and $v_3$, which are dominant contributions to this anisotropy \cite{Luzum:2013yya}. 
The measurement of these observables can further be categorized by ``centrality classes'', which describe the degree of overlap of the colliding nuclei at impact. Figure~\ref{fig:centrality} visualizes two collisions at low and high centrality. We make use of $v_2$ and $v_3$ in the 0--5\% centrality class (the 5\% of collisions with the most overlap of the colliding nuclei), and the 5--10\% class (the next 5\% of collisions with the most overlap), resulting in four observables of interest. This is a subset of the observables employed in a recent Bayesian analysis of the QGP \cite{everett2021phenomenological}, which yielded meaningful parameter constraints from experiments.

For inputs, the simulator has $d=17$ parameters: five describing the impact of the nuclei and the resulting energy deposition, two related to the transition of the systems into a quark-gluon plasma, nine for the temperature dependence of the shear and bulk viscosity of the quark-gluon plasma and the associated shear relaxation time, and one final parameter associated with the transition of the plasma's description from a fluid into a collection of individual particles. The same parametrization of this collision system was used in prior Bayesian analyses of the QGP (see \cite{liyanage2022efficient,ehlers2022bayesian,everett2022role}).

\subsection{Prediction and coverage}

Recall (see Section \ref{sec:qgpe}) that there are two key challenges for surrogate modeling of the QGP. First, prediction over this high-dimensional ($d=17$) space is challenging. Second, this is complicated by the \textit{costly} simulation runs needed to generate the training data, which requires thousands of CPU hours per design point. With limited supercomputing budget for this project, we can afford only $n^*=474$ training points, sampled from a Latin hypercube design \cite{mckay2000comparison}. Given such limited data in high dimensions, the task of accurate emulation with high predictive certainty is extremely challenging. We investigate whether, by learning and integrating dominant multi-physics structure for emulation, the AdMIn-GP can yield accurate and confident surrogate modeling in this challenging setting.

As before, we compared the proposed AdMIn-GP with the same existing models in our simulations (see Section \ref{sec:evaluation}). To evaluate predictive performance, we did not use an additional testing set due to the highly expensive nature of simulating such a test set. Rather, we made use of random splits of the full $n^*=474$ into training sets of sample sizes $n=200$, $250$, $300$ and $350$, with the remaining points used for testing. This random training-testing split is replicated 10 times to measure error variability. The same three metrics (RMSE, CRPS and empirical coverage rate) are used to gauge predictive performance.



\begin{figure}[!t]
\centering
\includegraphics[width=\textwidth]{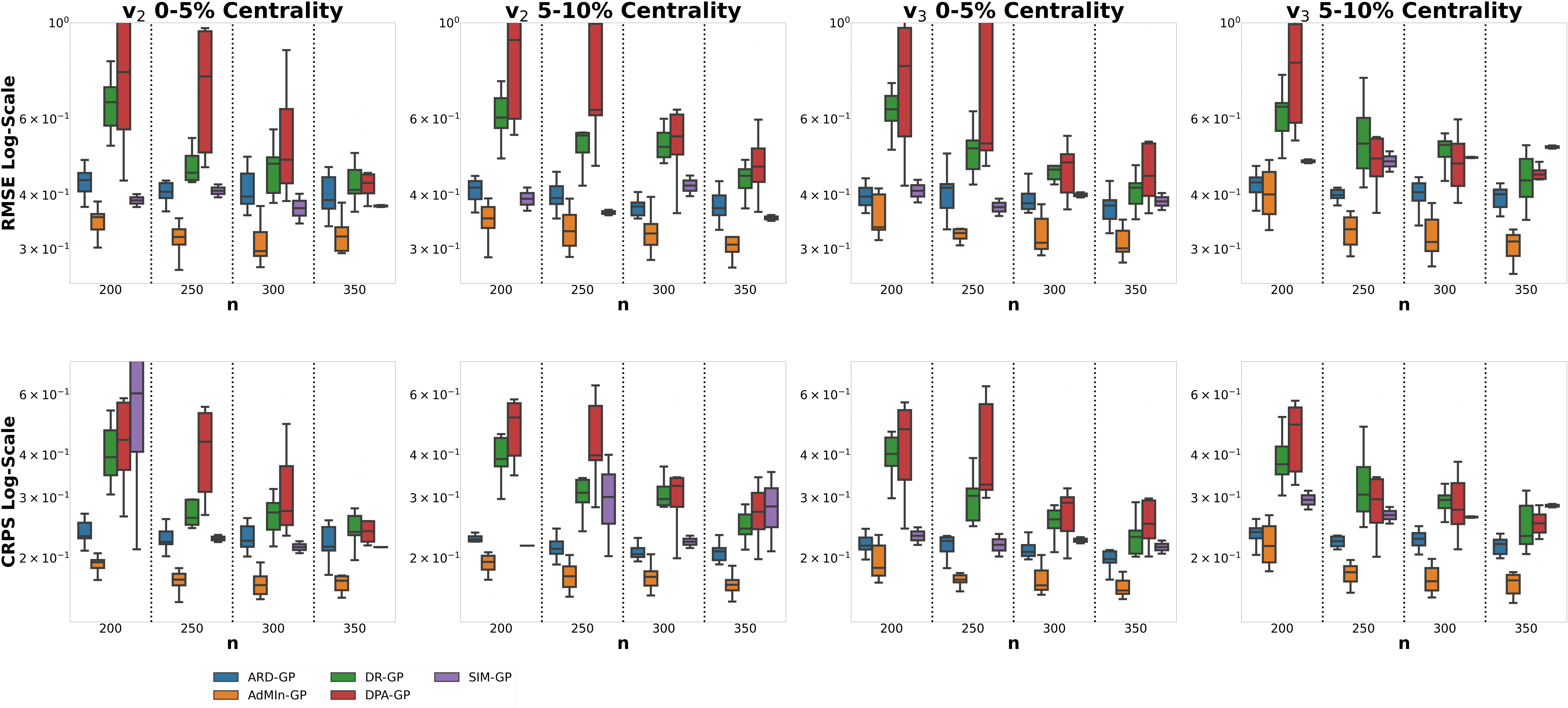}
\caption{Boxplots of predictive metrics (RMSE and CRPS) for the compared models over various sample sizes $n$. Each of the four columns corresponds to a different collision observable.}
\label{fig:real_results}
\end{figure}

Figure \ref{fig:real_results} shows the RMSEs and CRPSs of the compared surrogate models, over different sample sizes $n = 200, 250, 300$ and $350$ for each observable. We see that the proposed AdMIn-GP consistently yields significantly lower RMSEs to existing models, over all sample sizes $n$. Existing models that leverage some form of low-dimensional embedded structure (DR-GP, DPA-GP and SIM-GP) again yield mediocre predictions, with errors noticeably larger than the standard GP surrogate. From earlier simulations, this suggests that the underlying embedded structure may indeed be \textit{misspecified} by existing models; in retrospect, this is not surprising, as such models do not account for the presence of multi-physics. The standard GP surrogate, ARD-GP, yields relatively high errors over all observables, and does not appear to decrease as sample size $n$ increases. This confirms the limitations outlined in Section \ref{sec:gp}: standard GPs suffer from the curse-of-dimensionality of requiring an exponentially-growing sample size $n$ in dimension $d$ to achieve good predictive performance. Here, with small $n$ and large $d$, such models thus yield high errors that yield minimal decreases as $n$ increases. By eliciting and integrating the desired embedded structures from multi-physics for surrogate modeling, the AdMIn-GP can provide accurate predictions in this challenging small $n$, large $d$ setting. This suggests that our model can capture the true low-dimensional structure with much less misspecification. We will show later that this learned structure can further guide the extraction of interpretable multi-physics.

\begin{figure}[!t]
\centering
\includegraphics[width=\textwidth]{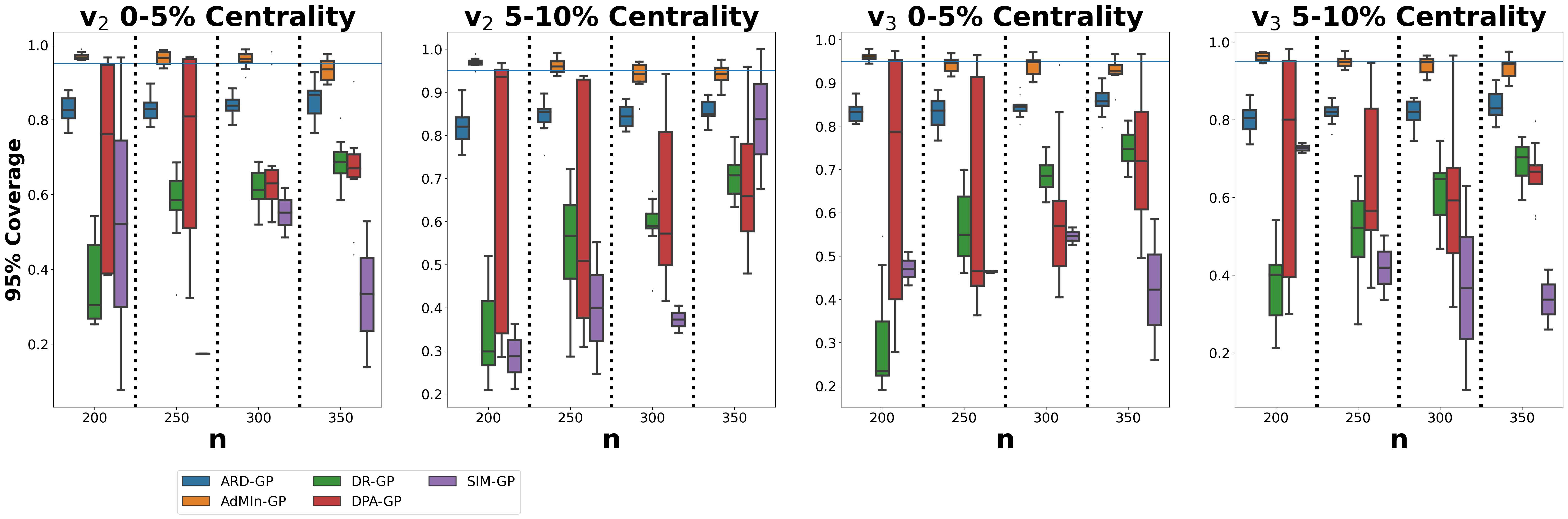}
\caption{Boxplots of the empirical coverage rates (of 95\% predictive intervals) for the compared models over various sample sizes $n$. Each column corresponds to a different collision observable.}
\label{fig:real_cov}
\end{figure}

Figure \ref{fig:real_cov} shows the corresponding empirical coverage rates of the compared models in this application, over different sample sizes $n$. As in simulations, we see that the proposed model provides much closer rates to the desired nominal 95\% rate for all observables and sample sizes. Existing models, however, yield much lower coverage rates than 95\%, particularly for models that attempt to fit embedded low-dimensional structure. There may be two reasons for this. First, as noted before, such existing models largely do not consider uncertainties with estimating the underlying embedding structure from data. With small sample sizes $n$ in high dimensions $d$, these uncertainties can be large and may result in significant undercoverage if ignored. Second, from earlier analysis, such existing models likely misspecified the more complex embedding structure arising from the multi-physics system. In doing so, the fitted model would be overconfident in predicting the more complex response surface, resulting in large undercoverage of predictive intervals.

\subsection{Extracting dominant multi-physics}
\label{sec:extract}

\begin{figure}[!t]
    \centering
    \includegraphics[width = \textwidth]{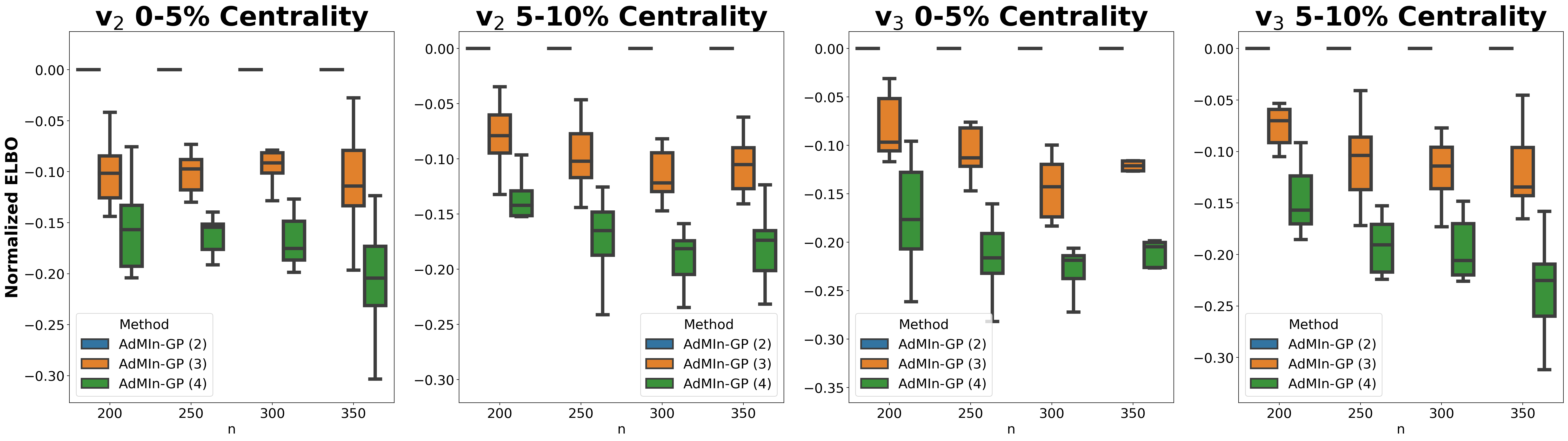} 
    \caption{Boxplots of the ``normalized'' ELBO bound \eqref{eq:elbo4} for the AdMIn-GP with $L$ additive components over various sample sizes $n$.. Negative values suggest that the ELBO model selection procedure prefers the $L=2$ component model.}
    \label{fig:mix_choose2}
\end{figure}

Finally, we explore the model and variable selection performance of the AdMIn-GP and discuss its implications for extracting multi-physics on the collision system. We first investigate the selected number of additive components $L$, then examine the selected important variables in the estimated embedding matrices, and how this might guide the extraction of interpretable multi-physics dominant within the QGP.


Consider first the selection of the number of additive components $L$ via the ELBO bound \eqref{eq:elbo4}. Here, we explored choices of $L$ from 2 to 4 components; the setting of $L=1$ is excluded here since the QGP is known to be influenced by \textit{multiple} distinct types of physics \cite{everett2021multisystem}. Figure \ref{fig:mix_choose2} shows the ``normalized'' ELBO bound, i.e., the ELBO bound \eqref{eq:elbo4} subtracted from the baseline ELBO bound with $L=2$, for different sample sizes $n$ and different observables. We see that, for all observables, the AdMIn-GP selects the baseline $L=2$ setting, which hints at two distinct dominant physics at play within the QGP. Further, as sample size $n$ increases, we see a larger gap in the ELBO bound between the baseline $L=2$ and larger choices of $L$, which again suggests the presence of two dominant physics.


\begin{figure}[!t]
    \centering
    \includegraphics[width=\textwidth]{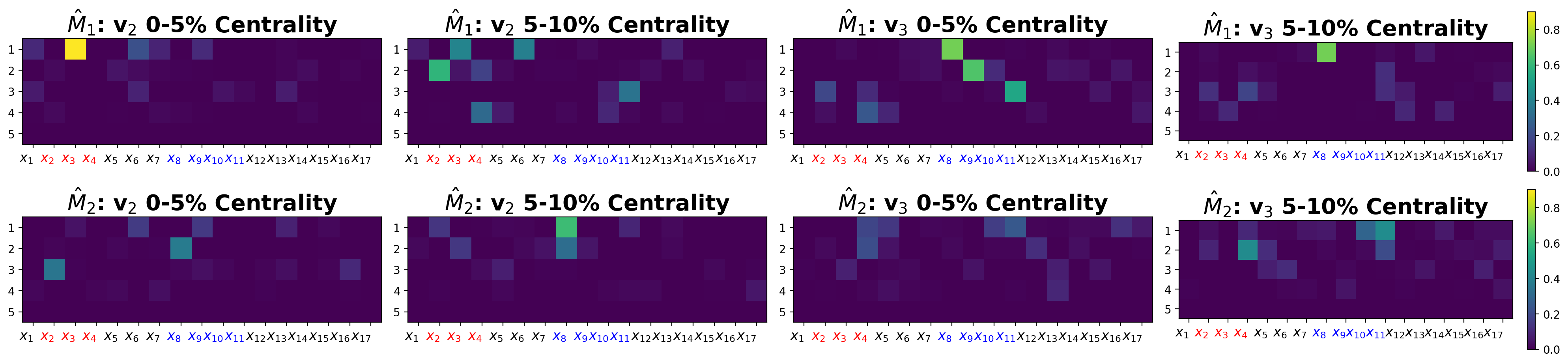} 
    \caption{Absolute values of the estimated matrices $\hat{\bm{M}}_1$ and $\hat{\bm{M}}_2$ for each of the four QGP observables. The first two columns are the estimated matrices for the first asymmetry observable at centrality $0-5\%$ and $5-10\%$, and the last two columns are for the second asymmetry observable at centrality $0-5\%$ and $5-10\%$. Parameters corresponding to collision initial conditions and plasma shear viscosity are highlighted in red and blue, respectively.}
    \label{fig:embed_qgp}
\end{figure}


The nature of this dominant multi-physics within the QGP can further be gleaned via a careful inspection of the estimated embedding matrices. Figure \ref{fig:embed_qgp} shows the estimated matrices $\hat{\bm{M}}_1$ and $\hat{\bm{M}}_2$ for the four observables. There are several illuminating observations to note. Consider the first observable $v_2$, i.e., the left two observables in Figure \ref{fig:embed_qgp}. We see that a majority of non-zero entries selected by the \textit{first} embedding matrix $\hat{\bm{M}}_1$ (top row) involves the parameters $x_2$, $x_3$ and $x_4$ (colored red on plot). Upon closer inspection, these three parameters (Trento-$p$, $\sigma_k$ and nucleon width) correspond specifically to the \textit{initial conditions} of the collision system~\cite{Moreland:2014oya}. It is well-accepted that the geometry and fluctuations of the collision initial conditions 
are directly related to momentum asymmetry \cite{Ollitrault:1992bk,miller2003eccentricity,PHOBOS:2006dbo,Alver:2010gr,Gardim:2011xv}, thus the first component of the AdMIn-GP likely captures the effect of such dominant initial conditions. Similarly, the \textit{second} embedding matrix (bottom row) selects non-zero entries on the parameters $x_8$ and, to a lesser extent, $x_9$ and $x_{11}$, which are three of the four parameters (colored blue in plot) used to parametrize the \textit{shear viscosity} of the plasma (see \cite{everett2021multisystem} for a detailed discussion of this parametrization). This dependence on shear viscosity affirms the sensitivity analysis in \cite{everett2021phenomenological}, which found a strong dependence of the observable $v_2$ on parameter $x_8$, but little dependence on the fourth shear viscosity parameter $x_{10}$. The second component of the AdMIn-GP thus likely captures the dependence of $v_2$ on plasma shear viscosity. One also observes a slight dependence on initial conditions (i.e., the parameters $x_2, x_3, x_4$) for the second matrix $\hat{\bm{M}}_2$, which confirms the known correlation between shear viscosity with the initial condition parameters of the QGP~\cite{Heinz:2013th,Gale:2013da}.

Consider next the second observable $v_3$, i.e., the right two observables in Figure \ref{fig:embed_qgp}. For the \textit{first} embedding matrix $\hat{\bm{M}}_1$, we notice a much larger dependence (compared to $v_2$) on the four shear viscosity parameters ($x_8-x_{11}$). This is not surprising from a physics perspective, as this asymmetric observable is related to shorter-scale structures of the plasma fluid that are damped more strongly by viscosity~\cite{Heinz:2013th,Gale:2013da}. This stronger dependence on the shear viscosity is now captured in $\hat{\bm{M}}_1$ as the dominant physics, with $\hat{\bm{M}}_2$ capturing a combination of initial condition parameters and their interaction with plasma shear viscosity. This extraction of \textit{interpretable} multi-physics in observables $v_2$ and $v_3$ not only affirms the modeled embedding structure in the AdMIn-GP, but also provides a data-driven approach for discovering and verifying dominant physical phenomena within the QGP.

Finally, we note that, for each of the two momentum asymmetry observables $v_2$ and $v_3$, the extracted dominant multi-physics (in the form of important variables in $\hat{\bm{M}}_1$ and $\hat{\bm{M}}_2$) are quite similar over different centralities; similar groups of parameters are selected as active or inert between the two centrality bins $0 - 5\%$ and $5 - 10\%$. This is consistent with the understanding that the centrality dependence of $v_2$ and $v_3$ is moderate~\cite{Heinz:2013th,Gale:2013da}, and similar results should be expected for neighboring centralities.


\section{Conclusion}
\label{sec:con}

The quark-gluon plasma (QGP) is a unique phase of nuclear matter, theoretized to have filled the Universe shortly after the Big Bang. The study of this plasma requires a probabilistic surrogate model, which can efficiently emulate the complex relationship between many physical parameters $\bm{x}$ and its observables $f(\bm{x})$ with expensive (and thus limited) training data. We thus propose a new Additive Multi-Index Gaussian process (AdMIn-GP) model, which features a flexible additive model of GPs with each component active on different latent low-dimensional linear embeddings. This structure is directly guided by prior knowledge that the QGP is controlled by several dominant multi-physics. We present an efficient framework for model fitting via a carefully constructed variational inference approach with inducing points. We then demonstrate the improved predictive performance and uncertainty quantification of the AdMIn-GP over competing methods in a suite of numerical experiments and for surrogate modeling of the QGP, and show it may facilitate the extraction of interpretable multi-physics for guiding scientific discoveries.

\if1\blind{
\noindent \textbf{Acknowledgements}: The authors gratefully acknowledge funding from NSF CSSI Frameworks grant 2004571 (KL, SM), NSF DMS 2210729, 2220496 (KL, SM) and U.S. Department of Energy Grant DE-FG02-05ER41367 (SAB, JFP). The authors also thank the JETSCAPE collaboration for useful comments and feedback.
}
\fi

\spacingset{1.0}
\bibliography{bib}

\pagebreak

\begin{appendices}
\label{appendix}

\section{Closed-form expressions for $\bs{\Psi}_1^l$ and $\bs{\Psi}_2^l$}
\label{appendix:integral}

We provide in the following a detailed derivation of the closed-form expectations for $\bs{\Psi}_1^l = [\phi_{1,i,j}^l]$ and $\bs{\Psi}_2^l = [\phi_{2,i,j}^l]$, which then provides tractable bounds for variational inference. The integrals to be computed are:
\begin{align}
\begin{split}
    &\phi_{1,i,j}^l = \int \nu_l \exp\left\{ -\frac{|| \bm{M}_l\bm{x}_i - \bm{z}_j||^2}{2} \right\} \pi_{\mathcal{MN}}(\bm{M}_l| \bm{A}_l, \bm{H}_l, \bm{U}_l) d\bm{M}_l \\
    &\phi_{2,i,j}^l =  \sum_{k=1}^n \int \nu_l^2 \exp\left\{ -\frac{||\bm{z}_i - \bm{M}_l\bm{x}_k||^2}{2} \right\} \exp\left\{ -\frac{||\bm{z}_j - \bm{M}_l\bm{x}_k||^2}{2} \right\} \pi_{\mathcal{MN}}(\bm{M}_l| \bm{A}_l, \bm{H}_l, \bm{U}_l) d\bm{M}_l,
\end{split}
\label{eq:int}
\end{align}
where $\pi_{\mathcal{MN}}(\cdot|\bm{A},\bm{H},\bm{U})$ denotes the density for the matrix normal distribution with mean matrix $\bm{A}$, row-wise covariance matrix $\bm{H}$, and column-wise covariance matrix $\bm{U}$.\\

We will derive these integrals first in the above general setting (i.e., with general forms for $\bm{H}_l$ and $\bm{U}_l$), then show how to simplify these expression under a simplified variational form with $\bm{H}_l = \bm{I}_{p \times p}$ and $\bm{U}_l = \bm{v}_l \bm{I}_{d \times d}$. For the general setting, first observe that, since $\bm{M}_l \sim \mathcal{MN}(\bm{A}_l, \bm{H}_l, \bm{U}_l)$, it follows that $\mbox{vec}(\bm{M}_l) \sim \mathcal{MVN}(\mbox{vec}(\bm{A}_l), \bm{U}_l \otimes \bm{H}_l)$. Thus, the required computation reduces to one involving a multivariate Gaussian integral. From the first expression in \eqref{eq:int}, we then have:
\begin{align*}
    \phi_{1,i,j}^l &=   C_j \int \exp\left\{-\frac{1}{2}\left(\text{tr}(\bm{M}_l^T\bm{M}_l\bm{x}_i\bm{x}_i^T) - 2\text{tr}(\bm{M}_l^T\bm{z}_j\bm{x}_i^T \right) \right\}  \\
    & \quad \times \exp\left\{-\frac{1}{2} \left( \mbox{vec}(\bm{M}_l)^T(\bm{U}_l \otimes \bm{H}_l)^{-1}\mbox{vec}(\bm{M}_l) - 2\mbox{vec}(\bm{M}_l)^T(\bm{U}_l \otimes \bm{H}_l)^{-1}\mbox{vec}(\bm{A}_l) \right) \right\} d\mbox{vec}(\bm{M}_l)
\end{align*}
where $C_j = \nu_j \cdot {\exp\Big(-\frac{1}{2} \mbox{vec}(\bm{A}_l)^T(\bm{U}_l\otimes \bm{H}_l)^{-1}\mbox{vec}(\bm{A}_l) - \frac{1}{2}\bm{z}_j^t\bm{z}_j \Big)}/((2\pi)^{\frac{pd}{2}}|\bm{U}_l \otimes \bm{H}_l|^{\frac{1}{2}})$. Recall that $\text{tr}(\bm{M}_l^T\bm{M}_l\bm{x}_i\bm{x}_i^T) = \mbox{vec}(\bm{M}_l)^T(\bm{x}_i\bm{x}_i^T \otimes \bm{I}_{d\times d}) \mbox{vec}(\bm{M}_l)$ and $\text{tr}(\bm{M}_l^T\bm{z}_j\bm{x}_i^T) = \mbox{vec}(\bm{M}_l)^T\mbox{vec}(\bm{x}_j\bm{x}_i^T)$. We can now complete the square, such that:
\begin{align}
\begin{split}
    \phi_{1,i,j}^l &= \nu \cdot \frac{\exp\left\{-\frac{1}{2} \mbox{vec}(\bm{A}_l)^T(\bm{U}_l\otimes \bm{H}_l)^{-1}\mbox{vec}(\bm{A}_l) - \frac{1}{2}\bm{z}_j^t\bm{z}_j \right\} }{(2\pi)^{\frac{pd}{2}}|\bm{U}_l \otimes \bm{H}_l|^{\frac{1}{2}} | (\bm{x}_i\bm{x}_i^T \otimes \bm{I}_{p \times p}) + (\bm{U}_l \otimes \bm{H}_l)^{-1}|^{\frac{1}{2}} } \\
    & \quad \times \exp\left\{\frac{1}{2}(\mbox{vec}(\bm{z}_j\bm{x}_i^T) + (\bm{U}_l \otimes \bm{H}_l)^{-1}\mbox{vec}(\bm{A}_l))^T \big((\bm{x}_i\bm{x}_i^T \otimes \bm{I}_{p \times p}) + (\bm{U}_l \otimes \bm{H}_l)^{-1}\big)^{-1} \right. \\
    & \left. \quad \times(\mbox{vec}(\bm{z}_j\bm{x}_i^T) + (\bm{U}_l \otimes \bm{H}_l)^{-1}\mbox{vec}(\bm{A}_l)) \right\}.
    \label{eq:int1}
    \end{split}
\end{align}
This provides a closed-form expression for computing the entries of $\bs{\Psi}_1^l$.\\

We can perform the exact same procedure for $\phi_{2,i,j}^l$ with little modification. From the second expression in \eqref{eq:int}, define the $k$-th component as:
\begin{align}
\begin{split}
    &\phi_{2,i,j,k}^l = \int \nu_l^2 \exp\left\{ -\frac{||\bm{z}_i - \bm{M}_l\bm{x}_k||^2}{2} \right\} \exp\left\{ -\frac{||\bm{z}_j - \bm{M}_l\bm{x}_k||^2}{2} \right\} \pi_{\mathcal{MN}}(\bm{M}_l| \bm{A}_l, \bm{H}_l, \bm{U}_l) d\bm{M}_l \\
    &=  \nu_l^2 \cdot \frac{\exp\left\{-\frac{1}{2} \mbox{vec}(\bm{A}_l)^T(\bm{U}_l \otimes \bm{H}_l)^{-1}\mbox{vec}(\bm{A}_l) - \frac{1}{2}(\bm{z}_j + \bm{z}_i)^T(\bm{z}_j + \bm{z}_i) \right\} }{(2\pi)^{\frac{pd}{2}}|\bm{U}_l \otimes \bm{H}_l|^{\frac{1}{2}} | (2\bm{x}_k\bm{x}_k^T \otimes \bm{I}_{p \times p}) + (\bm{U}_l \otimes \bm{H}_l)^{-1}|^{\frac{1}{2}} } \\
    &\times \exp\left\{\frac{1}{2}(\mbox{vec}((\bm{z}_j + \bm{z}_i)\bm{x}_k^T) + (\bm{U}_l \otimes \bm{H}_l )^{-1}\mbox{vec}(\bm{A}_l ))^T \big((2\bm{x}_k\bm{x}_k^T \otimes \bm{I}_{p \times p}) + (\bm{U}_l \otimes \bm{H}_l)^{-1}\big)^{-1} \right. \\
    & \left. \times (\mbox{vec}((\bm{z}_j + \bm{z}_i)\bm{x}_k^T) + (\bm{U}_l \otimes \bm{H}_l)^{-1}\mbox{vec}(\bm{A}_l)) \right\}.
    \label{eq:int2}
    \end{split}
\end{align}
Together with \eqref{eq:int}, this provides a closed-form expression for computing the entries of $\bs{\Psi}_2^l$. \\

One disadvantage with the above general setting (i.e., with general forms for $\bm{H}_l$ and $\bm{U}_l$) is that the closed-form expressions \eqref{eq:int1} and \eqref{eq:int2} require the inverting of $pd\times pd$ matrices, which requires $\mathcal{O}(p^3d^3)$ work and thus can be costly. We show next that we can greatly reduce this computational complexity if we restrict the variational form to $\bm{U}_l = \bm{v}_l \bm{I}_{d \times d}, \bm{H}_l = \bm{I}_{p \times p}$. From this restricted variational form, one can show that:
\begin{align*}
\phi_{1,i,j}^l &=   C_j \int \exp\left\{-\frac{1}{2}\big(\mbox{tr}(\bm{M}_l^T\bm{M}_l\bm{x}_i\bm{x}_i^T) - 2\mbox{tr}(\bm{M}_l^T\bm{z}_j\bm{x}_i^T \big) \big) \right\} \\
    &  \quad \times \exp\left\{-\frac{1}{2} \big( \mbox{tr}(\bm{M}_l^T\bm{M}_l\bm{U}_l^{-1}) - 2\mbox{tr}(\bm{M}_l^T\bm{A}_l\bm{U}_l^{-1}) \right\} d\bm{M}_l,
\end{align*}
where $C_j = {\exp\big(-\frac{1}{2}\mbox{tr}(\bm{U}_l^{-1}\bm{A}_l^T\bm{A}_l) - \frac{1}{2}\bm{z}_j \bm{z}_j^T \big) }/({|\bm{U}_l|^{\frac{d}{2}}2\pi^{\frac{pd}{2}}})$. We can then use the matrix-normal version of completing the square to obtain:
\begin{align}
    \phi_{1,i,j}^l = \frac{C_j}{|\bm{U}_l^{-1} + \bm{x}_i\bm{x}_i^T|^{\frac{d}{2}}} \exp\left\{\frac{1}{2}\mbox{tr}\big((\bm{z}_j\bm{x}_i^T + \bm{A}_l\bm{U}_l^{-1})^T(\bm{z}_j\bm{x}_i^T + \bm{A}_l\bm{U}_l^{-1})(\bm{U}_l^{-1} + \bm{x}_i\bm{x}_i^T)^{-1}\big)\right\}.
    \label{eq:int3}
\end{align}

The above closed-form expression for $\phi_{1,i,j}^l$ (under the restricted variational form) can be evaluated in significantly less work that the earlier $\mathcal{O}(p^3d^3)$. First, note that $C_j$ can be computed (for all $j$) in $\mathcal{O}(d^2p + mp)$ work. Next, we can make use of careful linear algebra simplifications to avoid expensive matrix inversions and multiplications in the exponent term in \eqref{eq:int3}. Note that the below expression admit rank-one updates:
\begin{align*}
(\bm{U}^{-1} + \bm{x}_i\bm{x}_i^T)^{-1}  = \bm{U} - \frac{\bm{U}\bm{x}_i\bm{x}_i^T\bm{U}}{1 + \bm{x}_i^T\bm{U}\bm{x}_i}, \quad |\bm{U}^{-1} + \bm{x}_i\bm{x}_i^T| = |\bm{U}^{-1}|(1 + \bm{x}_i^T\bm{U}\bm{x}_i).
\end{align*}
Thus, these quantities can be computed in $\mathcal{O}(d)$ because $\bm{U}$ is diagonal. Now examining the exponent term in \eqref{eq:int3}, we can distribute the product and re-arrange using the permutation invariance of $\text{tr}(\cdot)$ to obtain:
\begin{align*}
    &\mbox{tr}\big((\bm{z}_j\bm{x}_i^T + \bm{A}_l\bm{U}_l^{-1})^T(\bm{z}_j\bm{x}_i^T + \bm{A}_l\bm{U}_l^{-1})( \bm{U}_l - \frac{\bm{U}_l\bm{x}_i\bm{x}_i^T\bm{U}_l}{1 + \bm{x}_i^T\bm{U}_l\bm{x}_i})\big) \\
    &= \mbox{tr}(\bm{z}_j^T\bm{z}_j\bm{x}_i^T\bm{U}_l\bm{
    x}_i) - \frac{\mbox{tr}(\bm{z}_j^T\bm{z}_j\bm{x}_i^T\bm{U}_l\bm{x}_i\bm{x}_i^T\bm{U}_l\bm{x}_i)}{1 + \bm{x}_i^T\bm{U}_l\bm{x}_i} \\
    & \quad + 2\mbox{tr}(\bm{z}_j^T\bm{A}_l\bm{x}_i) - \frac{\mbox{tr}(2\bm{z}_j^T\bm{A}_l\bm{x}_i\bm{x}_i^T\bm{U}_l\bm{x}_i)}{1 +\bm{x}_i^T\bm{U}_l\bm{x}_i} \\
    & \quad + \mbox{tr}(\bm{U}_l^{-1}\bm{A}_l^T\bm{A}_l) - \frac{\mbox{tr}(\bm{x}_i^T\bm{A}_l^T\bm{A}_l\bm{x}_i)}{1 + \bm{x}_i^T\bm{U}_l\bm{x}_i}.
\end{align*} 
All these terms then decomposes into the product of scalars resulting from vector-matrix products. Combining these steps, we then see that $\bs{\Psi}_1^l$ can be computed in $\mathcal{O}(mnpd)$ work under the above restricted variational form. \\

The calculation for $\bs{\Psi}_2^l$ proceeds in an analogous manner.  We write
\begin{align*}
    \phi_{2,i,j,k}^l = \frac{C_j}{|\bm{U}_l^{-1} + 2\bm{x}_k\bm{x}_k^T|^{\frac{d}{2}}} \exp\left\{\frac{1}{2}\text{tr}\big(((\bm{z}_i +\bm{z}_j)\bm{x}_k^T + \bm{A}_l\bm{U}_l^{-1})^T( (\bm{z}_j + \bm{z}_i)\bm{x}_k^T + \bm{A}_l\bm{U}_l^{-1})(\bm{U}_l^{-1} + 2\bm{x}_k\bm{x}_k^T)^{-1}\big)\right\}.
\end{align*}
A similar decomposition (as in $\phi_{1,i,j}^l$) of the above exponent term can be performed, allowing $\bs{\Psi}_2^l$ to be computed in $\mathcal{O}(nm^2pd)$ work if $\bs{\Psi}_1^l$ has already been computed. 

\section{Derivation of ELBO bound}
\label{appendix:ELBO}

We show next the detailed derivation of the variational lower bound \eqref{eq:elbo4}. Recall that, for the AdMIn-GP, its generative process can be written as:
\begin{align*}
    &\bm{y} | \bm{f}_1 \dots \bm{f}_L  \sim \mathcal{MVN} \left(\sum_{l=1}^L \bm{f}_l, \beta^{-1}\bm{I}_{n \times n} \right), \\
    &\bm{f}_l | \bm{u}_l, \bm{M}_l \sim  \mathcal{MVN}( \bm{K}_{n,m}^{l} (\bm{K}_{m,m}^{l})^{-1} \bm{u}_l, \bm{K}_{n,n}^l + \bm{K}_{n,m}^l (\bm{K}_{m,m}^{l})^{-1}  \bm{K}_{m,n}^l) , \quad l = 1, \dots, L,\\
    &\bm{u}_l \sim \mathcal{MVN}(\bm{0}, (\bm{K}_{m,m}^{l})^{-1}) , \quad  l = 1, \dots, L,\\
    &\bm{M}_l \sim \mathcal{MN}(\bm{0}_{p \times d}, \bm{I}_{p \times p}, \bm{v}_l \bm{I}_{d \times d} ) , \quad l = 1, \dots, L.
\end{align*}

The variational approximation made is:
\begin{align*}
    p(\bm{f}_1 \dots \bm{f}_L, \bm{u}_1 \dots \bm{u}_L, \bm{M}_1 \dots \bm{M}_L|\bm{y}) \approx \phi(\bm{u})\prod_{l=1}^L p(\bm{f}_l|\bm{M}_l, \bm{u}_l)\phi(\bm{M}_l),
\end{align*}
where, as mentioned in the main paper, we choose variational distributions of the form:
\begin{align*}
&\phi(\bm{u})  \sim \mathcal{MVN}(\bm{u} | \bm{D}\bm{W}^{-1} \bar{\bm{y}}, \bm{D}\bm{W}^{-1}\bm{D}) \\
&\phi_l(\bm{M}_l) \sim \mathcal{MN}(\bm{A}_l, \bm{I}_{p \times p}, \bm{I}_{d \times d}\bm{v}_l).
\end{align*}

With this, we can then write out the ELBO bound, then simplify via the chosen variational distribution on $\phi(\bm{u})$:
\begin{align*}
    log \left\{p(\bm{y})\right\} &\geq \int \prod_{l=1}^L p(\bm{f}_l| \bm{u}_l, \bm{M}_l)\phi(\bm{u}_l)\phi(\bm{M}_l) \log\Big( \frac{p(\bm{y}|\bm{f}_1 \dots \bm{f}_L) \prod_{l=1}^L p(\bm{f}_l| \bm{u}_l, \bm{M}_l)p(\bm{u}_l)p(\bm{M}_l)}{\prod_{l=1}^L p(\bm{f}_l| \bm{u}_l, \bm{M}_l)\phi(\bm{u}_l)\phi(\bm{M}_l)} \Big) d\bm{f} d\bm{M} d\bm{u} \\
    &=  \int \phi(\bm{u})\phi(\bm{M}) \Big( \int \prod_{l=1}^L p(\bm{f}_l| \bm{u}_l, \bm{M}_l) \log\big( p(\bm{y}|\bm{f}_1 \dots \bm{f}_L) \big) d\bm{f} +  \log\left(\frac{p(\bm{u})}{\phi(\bm{u})}\right) \Big) d\bm{M} d\bm{u} \\
    & \quad \quad - \sum_{l= 1}^L KL(\phi(\bm{M}_l) ||p(\bm{M}_l)) \\
    &= \log\Bigg( \int p(\bm{u}) \exp\left\{ \mathbb{E}_{\phi(M)}\left[\log\big(N(\bm{y}|\sum_{l=1}^L \boldsymbol \alpha_l , \beta^{-1} \bm{I}_{n \times n})\big) \right] \right\} d\bm{u} \Bigg) \\
    & \quad \quad - \sum_{l=1}^L \frac{\beta}{2}\Big(\nu_l n - \mbox{tr}\big((\bm{K}_{m,m}^l)^{-1} \mathbb{E}_{\phi(\bm{M})}(\bm{K}_{m,n}^l\bm{K}_{n,m}^l) \Big) - \sum_{l=1} KL(\phi(\bm{M}_l) || p(\bm{M}_l)),
\end{align*}
where $\bs{\alpha}_l = \bm{K}_{n,m}^{l} (\bm{K}_{m,m}^{l})^{-1} \bm{u}_l$.\\

The key integral in the above expression is
\footnotesize
\begin{align*}
    &\int p(\bm{u}) \exp\left\{ \mathbb{E}_{\phi(\bm{M})}\left[\log\big(N(\bm{y}|\sum_{l=1}^L \boldsymbol \alpha_l , \beta^{-1} \bm{I}_{n \times n})\big) \right] \right\} d\bm{u}\\
    &= \left(\frac{\beta}{2\pi}\right)^{\frac{n}{2}} \exp\left\{-\frac{\beta}{2}\bm{y}^T\bm{y} \right\} \\
    & \times \int p(\bm{u})\exp\left\{-\frac{\beta}{2}\left( \sum_{l= 1}^L \sum_{s= 1}^L \bm{u}_l (\bm{K}_{m,m}^l)^{-1} \mathbb{E}_{\phi(\bm{M})}\left[ (\bm{K}_{m,n}^l\bm{K}_{n,m}^s)  \right] (\bm{K}_{m,m}^s)^{-1}\bm{u}_s  - 2\sum_{l=1}\bm{y}^T E_{\phi(\bm{M}_l)} \left[(\bm{K}_{n,m}^l) ( \bm{K}_{m,m}^l)^{-1}\bm{u}_l \right]  \right) \right\} d\bm{u} \\
    &= \left(\frac{\beta}{2\pi}\right)^{\frac{n}{2}} \exp\left(-\frac{\beta}{2}\bm{y}^T\bm{y} \right) \int p(\bm{u}) \exp\left(-\frac{\beta}{2} \bm{u}^T\bm{D}^{-1}\bm{P}\bm{D}^{-1}\bm{u} - \frac{\beta}{2} \bar{\bm{y}}^T\bm{D}^{-1}\bm{u}\right) d\bm{u}
\end{align*}
\normalsize
Finally, recalling that $p(\bm{u}) \sim \mathcal{MVN}(\bm{0}, \bm{D})$, this reduces to a simple multivariate Gaussian integral, and we can then write the final variational bound as:
\begin{align*}
    \log(p(\bm{y})) &\geq  \log\left( \frac{\beta^{\frac{n}{2}}\prod_{l=1}^L |\bm{K}_{m,m}^l|^{\frac{1}{2}}}{(2\pi)^{\frac{n}{2}}|\beta \bm{D} + \bm{P}|^{\frac{1}{2}}} \exp\left(\frac{1}{2}\beta^2\bar{\bm{y}}^T\bm{W}^{-1}\bar{\bm{y}})\right) \exp \left(-\frac{1}{2}\beta\bm{y}^T\bm{y} \right) \right) \\
    & \quad \quad - \sum_{l = 1}^L KL(\phi_l(\bm{M}_l) ||p(\bm{M}_l)) - \sum_{l=1}^L \bm{V}_l,
\end{align*}
where $\bm{V}_l = ({\beta}/{2}) (\nu_l n - \text{tr}\{(\bm{K}_{m,m}^l)^{-1} \boldsymbol{\Psi}_2^l\} )$ and $\bm{W} =  \beta \bm{P} + \bm{D}$.

\end{appendices}
\end{document}